\documentclass[amsmath,amssymb,ams, preprint, superscriptaddress,nofootinbib]{revtex4-2} 
\usepackage[a4paper,textwidth=17.9cm,textheight=26cm]{geometry}
\usepackage[english]{babel}
\usepackage{ulem}

\usepackage{siunitx} 

\addto\captionsenglish{}
\addto\captionsenglish{}

\usepackage{upgreek}
\usepackage{graphicx}
\renewcommand{\baselinestretch}{1.5} 

\usepackage[T1]{fontenc}
\usepackage{lmodern}
\usepackage{helvet}

\usepackage{lineno}

\usepackage{xr}
\makeatletter
\newcommand*{\addFileDependency}[1]{
  \typeout{(#1)}
  \@addtofilelist{#1}
  \IfFileExists{#1}{}{\typeout{No file #1.}}
}
\makeatother

\newcommand*{\myexternaldocument}[1]{
    \externaldocument{#1}
    \addFileDependency{#1.tex}
    \addFileDependency{#1.aux}
}
\myexternaldocument{Supplementary}

\begin{document}
\nolinenumbers

\title{All-photonic millimeter-wave and terahertz signal core at room and cryogenic temperatures}

\author{Jiawen Liu}
\altaffiliation{jiawen.liu@epfl.ch}
\affiliation{Hybrid Photonics Laboratory, École Polytechnique Fédérale de Lausanne (EPFL), CH-1015, Switzerland}
\affiliation{Center for Quantum Science and Engineering (EPFL), CH-1015, Switzerland}

\author{Chengli Wang}
\affiliation{National Key Laboratory of Materials for Integrated Circuits, Shanghai Institute of Microsystem and Information Technology, Chinese Academy of Sciences, Shanghai, China}

\author{Aleksei Gaier}
\affiliation{Hybrid Photonics Laboratory, École Polytechnique Fédérale de Lausanne (EPFL), CH-1015, Switzerland}
\affiliation{Center for Quantum Science and Engineering (EPFL), CH-1015, Switzerland}

\author{Tianyi Zhang}
\affiliation{Hybrid Photonics Laboratory, École Polytechnique Fédérale de Lausanne (EPFL), CH-1015, Switzerland}
\affiliation{Center for Quantum Science and Engineering (EPFL), CH-1015, Switzerland}

\author{Gabriel Jülg}
\affiliation{Hybrid Photonics Laboratory, École Polytechnique Fédérale de Lausanne (EPFL), CH-1015, Switzerland}
\affiliation{Center for Quantum Science and Engineering (EPFL), CH-1015, Switzerland}

\author{Yazan Lampert}
\affiliation{Hybrid Photonics Laboratory, École Polytechnique Fédérale de Lausanne (EPFL), CH-1015, Switzerland}
\affiliation{Center for Quantum Science and Engineering (EPFL), CH-1015, Switzerland}

\author{André Garcia Primo}
\affiliation{Hybrid Photonics Laboratory, École Polytechnique Fédérale de Lausanne (EPFL), CH-1015, Switzerland}
\affiliation{Center for Quantum Science and Engineering (EPFL), CH-1015, Switzerland}

\author{Alessandro Tomasino}
\affiliation{Hybrid Photonics Laboratory, École Polytechnique Fédérale de Lausanne (EPFL), CH-1015, Switzerland}
\affiliation{Center for Quantum Science and Engineering (EPFL), CH-1015, Switzerland}

\author{Hua Li}
\affiliation{National Key Laboratory of Materials for Integrated Circuits, Shanghai Institute of Microsystem and Information Technology, Chinese Academy of Sciences, Shanghai, China}

\author{Xin Ou}
\altaffiliation{ouxin@mail.sim.ac.cn}
\affiliation{National Key Laboratory of Materials for Integrated Circuits, Shanghai Institute of Microsystem and Information Technology, Chinese Academy of Sciences, Shanghai, China}
 
\author{Ileana-Cristina Benea-Chelmus}
\altaffiliation{cristina.benea@epfl.ch}
\affiliation{Hybrid Photonics Laboratory, École Polytechnique Fédérale de Lausanne (EPFL), CH-1015, Switzerland}
\affiliation{Center for Quantum Science and Engineering (EPFL), CH-1015, Switzerland} 
\date{\today}

\begin{abstract}

Millimeter-wave and terahertz technologies offer new opportunities for high-capacity communications, sensing, and quantum systems, but electronic signal chains above 100~GHz are limited by weak sources, noisy receivers, and lossy interconnects. Photonic approaches mitigate these constraints by enabling chip-scale high-frequency components, yet existing implementations generally retain lossy off-chip electrical or free-space links, hindering dense and compact integration for cryogenic or space-constrained settings. Here, we show a monolithic lithium tantalate photonic chip, termed a signal core, which generates, modulates, routes, and detects continuous-wave signals at discrete frequencies between 100~GHz and 2.5~THz. The architecture integrates high-frequency functional modules on chip and uses exclusively fiber-based optical interfaces. Bidirectional coherent electro-optic conversion enables signal generation, detection, and nanosecond-scale amplitude and phase modulation, while micro-coplanar striplines interconnect the modules into an end-to-end signal chain on a single chip. Our modules achieve single-photon electro-optic coupling rates of $g_0/2\pi\approx10$~kHz and peak drive-power-normalized photon-flux conversion efficiencies of $1.0\%\mathrm{mW}^{-1}$ for generation and $0.5\%\mathrm{mW}^{-1}$ for detection. It also operates at 4~K in a dilution refrigerator through the same fiber-only optical links. Together, these capabilities establish an all-photonic platform for millimeter-wave-to-terahertz systems and offer a broadly applicable architecture for applications constrained by electrical-interconnect loss, heat load, and footprint.

\end{abstract}

\maketitle

\section*{Introduction}

\begin{figure*}[h!]
    \centering
    \includegraphics[width=1\textwidth]{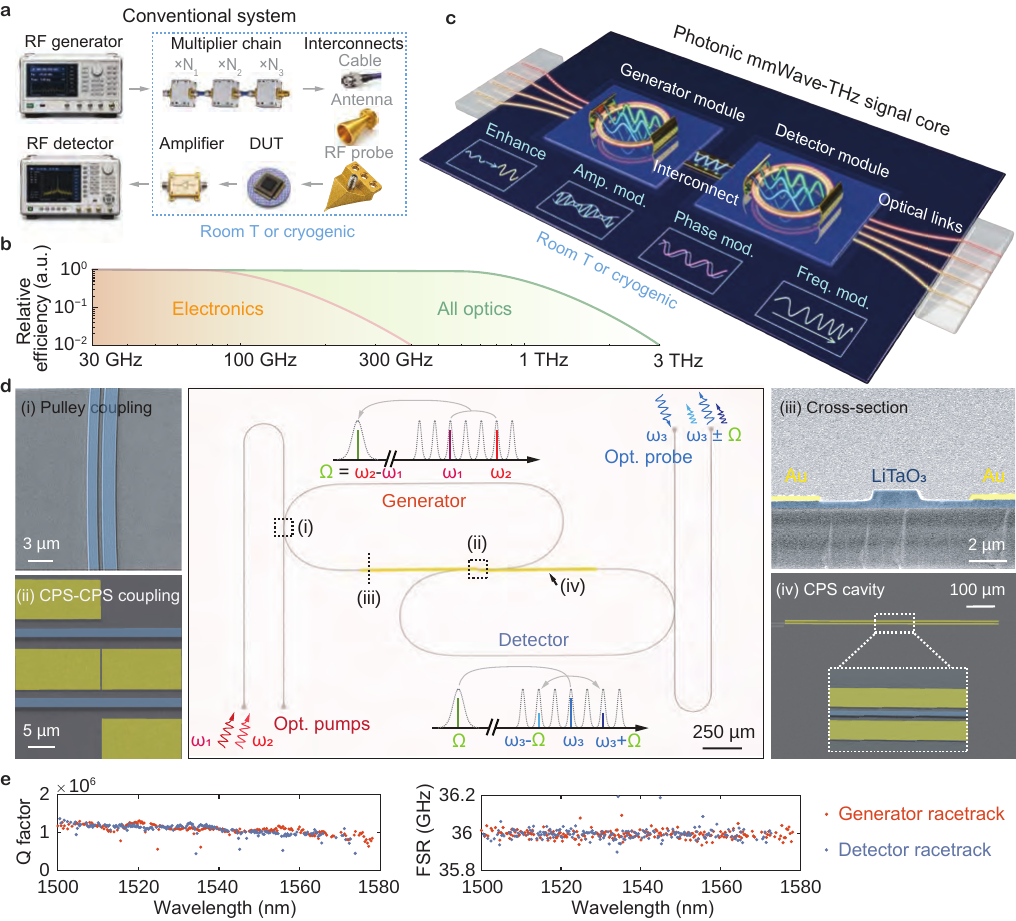}
    \renewcommand{\baselinestretch}{1} 
    \caption{\textbf{Photonic signal core for mmWave-THz frequencies.} 
    \textbf{a}. Conventional electronic signal chain using separate components for signal generation, transfer, and detection.
    \textbf{b}. Conceptual comparison of the spectral coverage accessible using conventional electronic and all-photonic systems.
    \textbf{c}. Photonic signal-core architecture integrating a generator, an on-chip interconnect and a detector on a single chip. The architecture provides resonant field enhancement and optical control of the amplitude, phase and frequency of the mmWave-THz signal. 
    \textbf{d}. Optical micrograph (main panel) and false-color scanning electron micrographs (insets) of the fabricated signal core. The architecture comprises generator and detector modules linked by capacitively coupled CPS cavities. Each module couples an optical racetrack resonator to the CPS cavity supporting the mmWave-THz field at angular frequency $\Omega$. Optical pumps at $\omega_1$ and $\omega_2$ generate $\Omega=\omega_2-\omega_1$, which is detected through sidebands at $\omega_3\pm\Omega$. Insets show (i) the bus-racetrack pulley coupler; (ii) the CPS coupling region with a nominal 200~nm gap; the racetracks are separated by 12~\textmu m to ensure negligible optical coupling; (iii) cross-sections of the lithium tantalate waveguide and gold CPS; and (iv) the aligned CPS and racetrack waveguides for electro-optic interaction.
    \textbf{e}. Optical characterization of the generator (red) and detector (blue) resonators, showing loaded quality factors of approximately $1.2\times10^6$ and matched free spectral ranges near 36~GHz over an 80-nm (approximately 10-THz) span.
    }
    \label{Fig1}
\end{figure*}

Millimeter-wave (mmWave) and terahertz (THz) frequencies occupy a technologically challenging region between microwave electronics and optics. Access to this spectrum enables high-capacity wireless communications and data links~\cite{koenig2013wireless,nagatsuma2016advances,dang2020should,cai2026surpassing,zhang2026integrated}, sensing and spectroscopy~\cite{liebermeister2021optoelectronic,jiang2024terahertz,djevahirdjian2023frequency,li2024plasmonic}, and interfaces to quantum systems~\cite{weaver2024integrated,anferov2025millimeter,warner2025coherent,multani2026integrated}. However, above 100~GHz, conventional electrical signal chains become increasingly difficult to integrate and scale (Fig.~\ref{Fig1}a)~\cite{leitenstorfer20232023,gruber2025sub}. Generation commonly relies on cascaded frequency multiplication: an overall multiplication factor $N$ introduces an ideal phase-noise penalty of $20\log_{10}N$~dB, while finite conversion efficiency at each stage increases the power loss~\cite{siles2018new,tao2025ultrabroadband}. Additional attenuation in cables, probes, packages, and free-space couplers requires signal amplification and ancillary hardware, increasing power consumption, system complexity, and heat load.

Photonics and optoelectronics provide alternative routes to this spectral regime. Photomixing in high-speed photodiodes or photoconductive antennas produces mmWave and THz radiation, whereas optical heterodyne reception enables coherent detection~\cite{nellen2020experimental,jia2022integrated,zhao2026terahertz,lepeshov2017enhancement,deumer2025advancing}. Quantum cascade lasers~\cite{kohler2002terahertz,khalatpour2021high} and quantum-well THz detectors~\cite{jeannin2020high} offer compact direct sources and receivers, respectively. However, these technologies generally implement individual functions and rely on free-space or off-chip links to form a complete system. Some source technologies still require substantial electrical power and active cooling, complicating their use in environments with stringent power and thermal budgets. Recent advances in photonic integration have combined wireless functions at frequencies up to 115~GHz~\cite{tao2025ultrabroadband}, demonstrated monolithic continuous-wave generation and modulation up to 500~GHz in lithium niobate devices~\cite{zhang2025monolithic}, and extended coherent emission at discrete resonances from 330~GHz to 2.3~THz through photomixing~\cite{wang2022coherent}. Collectively, these advances either combine multiple functions over a limited frequency range, or extend into the THz regime for selected functions. The remaining challenge is to realize a chip-scale, end-to-end continuous-wave signal chain covering the THz gap, where existing systems encounter severe bandwidth and efficiency constraints (Fig.~\ref{Fig1}b).

Electro-optic photonic circuits offer a compelling route to this goal. In these circuits, second-order nonlinear interactions mediated by $\chi^{(2)}$ enable a coherent, bidirectional conversion between optical and mmWave-THz fields without relying on charge transport or bandgap-engineered active media. This mechanism supports room-temperature generation and detection~\cite{herter2023terahertz,lampert2025photonics,tomasino2026large}, as well as modulation extending into the THz regime~\cite{ummethala2019thz,horst2025ultra}. Although electro-optic coupling is weak at the single-photon level~\cite{fan2018Superconducting,holzgrafe2020cavity}, it can be substantially enhanced in triply resonant configurations, in which two optical modes and one mmWave-THz mode are resonantly coupled~\cite{soltani2017efficient,mckenna2020cryogenic}. Moreover, for a given geometry and modal overlap, the zero-point electric field, and hence the single-photon electro-optic coupling rate, increases approximately with the square root of the frequency~\cite{zhang2025integrated,gaier2025wireless}. Together, cavity enhancement and favorable frequency scaling provide the physical basis for a monolithic signal core comprising generator and detector modules, on-chip high-frequency interconnects, and fiber-coupled interfaces for optical drive and readout (Fig.~\ref{Fig1}c). This optically interfaced architecture removes external high-frequency links, thereby reducing interconnect loss, packaging complexity, and cryogenic heat load.

Here we realize this signal core in thin-film lithium tantalate (Fig.~\ref{Fig1}d), a platform combining strong $\chi^{(2)}$ nonlinearity, low optical loss, a high optical-damage threshold, engineered phase matching, and wafer-scale manufacturability~\cite{wang2024lithium}. The generator and detector modules each couple a racetrack optical resonator to a coplanar-stripline (CPS) cavity, forming triply resonant converters for optical-to-mmWave-THz downconversion and reciprocal upconversion. Capacitively coupled CPS structures connect the modules for coordinated multifunctional operations and confine the high-frequency signal path within the chip to minimize losses. At room temperature, we demonstrate end-to-end signal chain operation at selected cavity resonances between 100~GHz and 2.5~THz. Across this range, the increasing single-photon electro-optic coupling competes with the rising mmWave-THz loss, revealing a frequency-dependent trade-off that governs system performance. Our signal core achieves strong single-photon electro-optic coupling rates of $g_0/2\pi\approx10$~kHz at frequencies between 180 and 900~GHz, with maximum pump-normalized photon-flux conversion efficiencies of $1.0\%/\mathrm{mW}$ for downconversion and $0.5\%/\mathrm{mW}$ for upconversion. 
Modulation of the optical pumps further enables coherent waveform synthesis with nanosecond-scale control over the amplitude, phase, and instantaneous frequency. At 4~K in a dilution refrigerator, we demonstrate the end-to-end operation at frequencies up to 576~GHz, through the same fiber-only interfaces, whereas the electronic-based signal chains would require complex multi-component assemblies or are currently difficult to access the covered frequency range.

The combination of wide frequency coverage, coherent waveform control, and cryogenic compatibility establishes a compact photonic platform for environments constrained by wiring density, heat load, and footprint. These capabilities are particularly relevant to emerging technologies such as mmWave superconducting qubits~\cite{anferov2025millimeter}, sub-THz transducers~\cite{multani2026integrated}, and superconducting traveling-wave parametric amplifiers being extended from microwave to mmWave operation~\cite{qiu2023broadband,tan2024operation}. More broadly, the architectural advance in this work offers a route to high-frequency systems that scale through photonic integration, rather than increasingly complex external interconnects.

\section*{Results}

\subsection*{1. Simultaneous triply resonant conversion}

The photonic signal core integrates two resonantly enhanced modules for mmWave-THz generation and detection (Fig.~\ref{Fig1}d). In the generator, optical pumps at angular frequencies $\omega_{1}$ and $\omega_{2}$ drive difference-frequency generation (DFG) in a lithium tantalate racetrack resonator, producing a field at $\Omega=\omega_{2}-\omega_{1}$ in its coplanar-stripline (CPS) cavity. A capacitive interconnect couples this field into the detector CPS cavity, where it electro-optically modulates a probe at $\omega_3$ and produces sidebands at $\omega_3\pm\Omega$.

End-to-end resonant enhancement requires both modules to satisfy the triply resonant condition at the same $\Omega$, which is more stringent than achieving it in either converter separately. This imposes four spectral requirements. First, $\omega_1$, $\omega_2$, and $\omega_3$ must align with resonances of their respective optical racetracks. Second, $\Omega$ must match the generator CPS resonance, while $\omega_3\pm\Omega$ must align with detector-racetrack modes. Third, the generator and detector CPS resonances must coincide so that the field can be resonantly transferred between modules. Finally, the optical free spectral ranges (FSRs) of the two racetracks must remain closely matched to preserve these relationships over the operating range.

We satisfy these conditions using nominally identical optical racetracks and CPS cavities in the two modules. Both racetracks exhibit loaded optical quality factors of approximately $1.2\times10^6$ (Fig.~\ref{Fig1}e), corresponding to linewidths near 160~MHz. Their measured FSRs remain close to 36~GHz over an 80~nm wavelength span (approximately 10~THz), with standard deviations below 25~MHz, or 0.07\% of the mean FSR. This uniformity allows the same mmWave-THz resonance at $\Omega$ to be addressed by both modules over multiple optical mode orders, extending resonantly enhanced conversion from an individual device to a multi-module signal chain. Detailed optical-resonator characterization is provided in Supplementary Information.

In our signal core, not every CPS mode matches an integer multiple of the 36~GHz optical FSR. For example, the fundamental CPS resonance at $\Omega/2\pi\approx90$~GHz corresponds to 2.5 FSRs and is therefore maximally offset from an integer-FSR separation. By contrast, higher-order resonances near 180, 360, and 540~GHz coincide with separations of five, ten, and fifteen FSRs, respectively, allowing both modules to become triply resonant simultaneously. Comparing these matched and mismatched modes provides a direct test of triply resonant enhancement in the mmWave-THz regime.

Spectral alignment alone does not ensure efficient resonant DFG. As the combined optical power increases, both pump tones must remain coupled to their respective generator-racetrack modes despite photorefractive and thermo-optical resonance shifts. We quantify simultaneous pump loading by $\eta_{\mathrm{load}}$, defined as the fraction of the combined input power coupled into the two resonant modes (Supplementary Information). Our thin-film lithium tantalate resonators maintain $\eta_{\mathrm{load}}>95\%$ even when the combined intracavity pump powers exceed 70~mW. By comparison, lithium niobate control devices matched in geometry, coupling regime, and loaded optical quality factor exhibit $\eta_{\mathrm{load}}<80\%$ under otherwise identical conditions. This difference is consistent with the reduced photorefractive resonance drift reported for thin-film lithium tantalate~\cite{powell2024dc,wang2024lithium}. The resulting dual-mode stability allows efficient frequency conversion without active resonance tracking, making our lithium tantalate implementation particularly well suited for stable, resonantly enhanced operation of the signal core.


\subsection*{2. Resonant on-chip mmWave-THz transfer}

\begin{figure}[h!]
    \centering
    \includegraphics[width=1\textwidth]{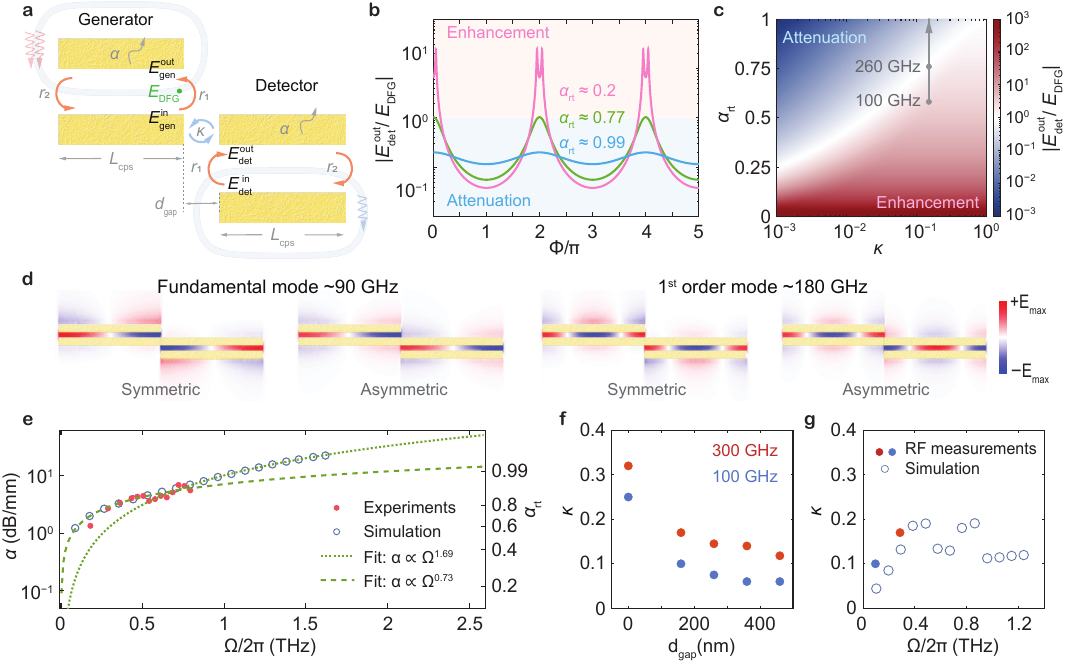}
    \renewcommand{\baselinestretch}{1} 
    \caption{\textbf{Coupled mmWave-THz CPS structures.} 
    \textbf{a}. Schematic of the coupled CPS cavities with internal electro-optic drive and readout. Each CPS is described by a power-attenuation coefficient $\alpha$, an inter-cavity field-coupling coefficient $\kappa$, and field-reflection coefficients $r_1$ and $r_2$ at the indicated boundaries.
    \textbf{b}. Calculated amplitude-transfer factor $|E_\mathrm{det}^\mathrm{out}/E_\mathrm{DFG}|$ for representative combinations of coupling coefficients $\kappa$ and round-trip attenuation factor $\alpha_\mathrm{rt}$, showing resonant enhancement and attenuation regimes.
    \textbf{c}. Amplitude-transfer factor $\left|E_{\mathrm{det}}^{\mathrm{out}}/E_{\mathrm{DFG}}\right|$ as a function of $\kappa$ and $\alpha_{\mathrm{rt}}$. Blue and red regions denote attenuation $\left(\left|E_{\mathrm{det}}^{\mathrm{out}}/E_{\mathrm{DFG}}\right|<1\right)$ and resonant enhancement $\left(\left|E_{\mathrm{det}}^{\mathrm{out}}/E_{\mathrm{DFG}}\right|>1\right)$, respectively. The gray line traces the operating trajectory of the present signal core as the frequency increases from 90~GHz to 1.6~THz. For the present device parameters, the trajectory remains in the resonant-enhancement regime below approximately 0.2~THz.
    \textbf{d}. Simulated electric-field distributions of the hybridized first and second longitudinal CPS modes near 90 and 180~GHz, respectively.
    \textbf{e}. Frequency-dependent power-attenuation coefficient $\alpha$ and corresponding round-trip field-attenuation factor $\alpha_{\mathrm{rt}}$. Red dots show values extracted in situ from the signal core, and blue circles show numerical simulations. Power-law fits yield $\alpha\propto\Omega^{0.73}$ below 700~GHz and $\alpha\propto\Omega^{1.69}$ above 700~GHz.
    \textbf{f}. Inter-cavity field-coupling coefficient $\kappa$ as a function of the capacitive gap $d_{\mathrm{gap}}$, measured using antenna-coupled test devices and a calibrated mmWave source near 100~GHz (blue dots) and 300~GHz (red dots).
    \textbf{g}. Frequency-dependent coupling coefficient $\kappa$ obtained from calibrated RF measurements (dots) and full-structure numerical simulations (circles).
    }
    \label{Fig2}
\end{figure}

A central feature of the signal-core architecture is the capacitive coupling between the CPS cavities of the generator and detector modules (Fig.~\ref{Fig2}a). Unlike conventional coupled resonators accessed through external electrical ports, these cavities are excited and read out internally through nonlinear frequency conversion. To quantify the field-transfer efficiency between modules, we represent the single-pass mmWave-THz field generated along the DFG interaction region as an effective lumped intracavity source of amplitude $E_{\mathrm{DFG}}$ at the coupling interface. The amplitudes $E_{\mathrm{gen}}^{\mathrm{in/out}}$ and $E_{\mathrm{det}}^{\mathrm{in/out}}$ denote the built-up fields entering and leaving this interface on the generator and detector sides, respectively (Fig.~\ref{Fig2}a). For a CPS of length $L_{\mathrm{cps}}$ and a power-attenuation coefficient $\alpha(\Omega)$, the single-pass field-transmission coefficient is $a=e^{-\frac{\alpha}{2}L_{\mathrm{cps}}}$. The generator-side intracavity field and detector-side intracavity field are then 
\begin{equation}
E_\mathrm{gen}^\mathrm{in} =
\frac{1 - r_1 r_2 a^2 e^{i\phi}}
{1 - 2r_1 r_2 a^2 e^{i\phi}
+ r_2^2 a^4 e^{2i\phi}}E_{\mathrm{DFG}}
\qquad
E_\mathrm{det}^\mathrm{out} =
\frac{i \kappa}
{1 - 2r_2r_1 a^2 e^{i\phi}+ r_2^2 a^4 e^{i2\phi}}
E_\mathrm{DFG}
\label{Eq_1}
\end{equation}
where $r_1$ and $r_2$ are the field-reflection coefficients at the boundaries identified in Fig.~\ref{Fig2}a, $\kappa$ is the inter-cavity field-coupling coefficient, and $\phi$ is the round-trip propagation phase (Supplementary Information).

In particular, the amplitude-transfer factor $|E_\mathrm{det}^\mathrm{out}/E_\mathrm{DFG}|$ can exceed unity, which means that resonant buildup in the coupled cavities can compensate for the low single-photon coupling rate. This regime occurs near the coupled-cavity eigenfrequencies when the round-trip field-attenuation factor $\alpha_\mathrm{rt}=\sqrt{1-a^2}$ is relatively low, as illustrated in Fig.~\ref{Fig2}b. This coupling also hybridizes nominally degenerate longitudinal modes into symmetric and antisymmetric supermodes, producing a resolvable resonance doublet when the mode splitting exceeds the loaded linewidths. Simulated electric-field profiles illustrate this hybridization for the first and second longitudinal CPS modes near 90 and 180~GHz, respectively (Fig.~\ref{Fig2}d).

We map the maximum transfer factor $|E_\mathrm{det}^\mathrm{out}/E_\mathrm{DFG}|$ as a function of $\kappa$ and $\alpha_{\mathrm{rt}}$, delineating the resonant-enhancement and attenuation regimes (Fig.~\ref{Fig2}c). To locate our fabricated signal core on this map, we measure $\alpha(\Omega)$ in situ using a previously developed electro-optic modulation method~\cite{gaier2025wireless} (Supplementary Information). The measurements agree with numerical simulations across the investigated frequency range (Fig.~\ref{Fig2}e). Power-law fits yield $\alpha\propto\Omega^{0.73}$ for $\Omega/2\pi<700$~GHz and $\alpha\propto\Omega^{1.69}$ at higher frequencies. The increase from an exponent near the $\Omega^{0.5}$ scaling associated with conductor loss toward the $\Omega^3$ dependence expected for radiation loss indicates a growing radiative contribution at high frequencies~\cite{keil2002high,frankel2002terahertz}.

The coupling coefficient $\kappa$ is governed primarily by the gap $d_{\mathrm{gap}}$ between the CPS structures. We extract $\kappa$ from the coupled-cavity response of test devices equipped with on-chip antennas at the CPS input (Supplementary Information), which are excited with a calibrated frequency-multiplier source to reduce uncertainty in incident mmWave power. Measurements across a range of gap sizes show that $\kappa$ decreases with increasing $d_{\mathrm{gap}}$, as expected for capacitive coupling (Fig.~\ref{Fig2}f). The nominal zero-gap contact provides an upper-bound benchmark of $\kappa\approx0.3$, whereas the 200-nm gap used in the signal core yields $\kappa\approx0.15$. Full-structure simulations extend this characterization beyond the frequency range accessible to the multiplier source, as shown in Fig.~\ref{Fig2}g and Supplementary Information.

Substituting the measured $\alpha$ and $\kappa$ into Equation~(\ref{Eq_1}) places the signal core in the resonant-enhancement regime up to $\Omega/2\pi\approx260$~GHz (Fig.~\ref{Fig2}c). Thus, in contrast to non-resonant transmission lines, our coupled-cavity link uses field buildup to increase the field delivered to the detector. Together with keeping the entire high-frequency path on chip and therefore avoiding losses associated with external interconnects, the signal-core architecture remains efficient across a broadband frequency range.


\subsection*{3. End-to-end operation up to 2.5~THz}

\begin{figure*}[h!]
    \centering
    \includegraphics[width=1\textwidth]{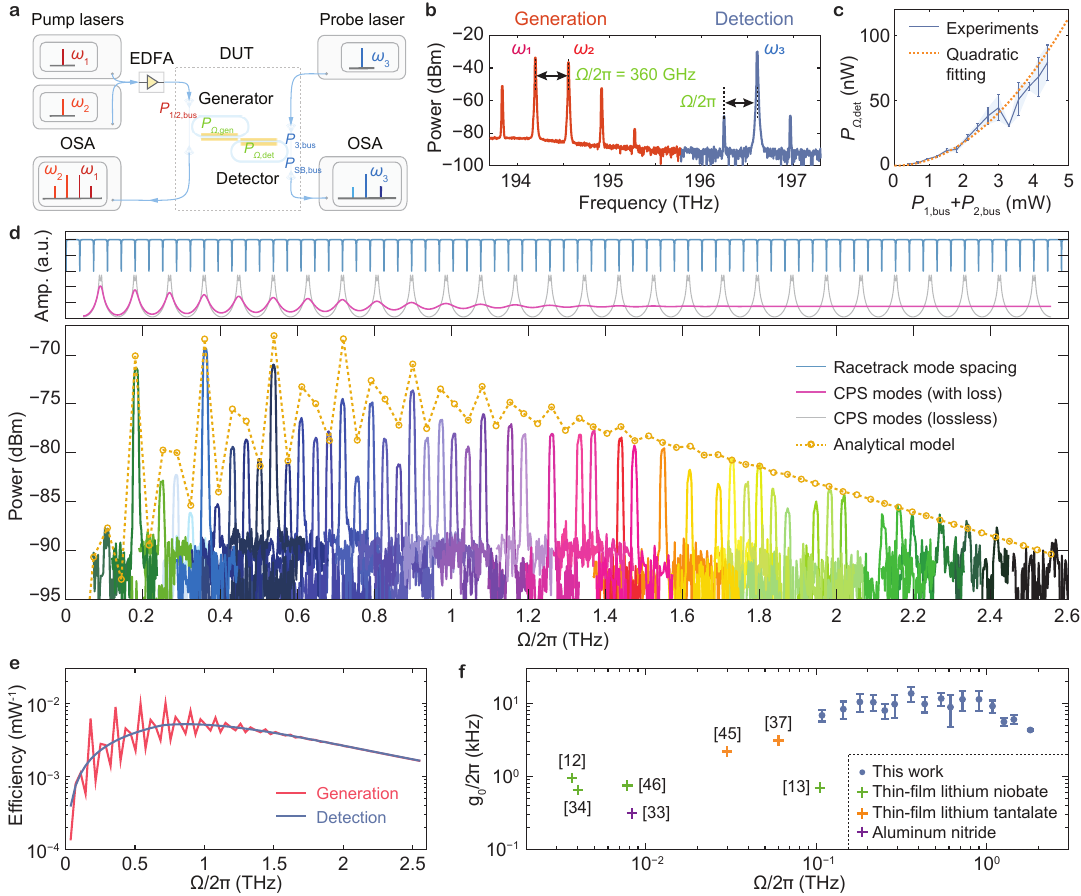}
    \renewcommand{\baselinestretch}{1} 
    \caption{\textbf{Monolithic mmWave-THz generation, on-chip transfer and detection.} 
    \textbf{a}. Schematic of the room-temperature experimental setup for mmWave-THz generation and detection. EDFA, erbium-doped fiber amplifier; DUT, device under test; OSA, optical spectrum analyzer. Blue lines denote optical-fiber connections.
    \textbf{b}. Representative optical spectra measured after the generator (red) and detector (blue) at $\Omega/2\pi=360$~GHz. Sidebands at $\omega_3 \pm \Omega$ arise from electro-optic upconversion of the mmWave-THz field generated and transferred on chip. Spectral features at the generator output are attributed to combined $\chi^{(2)}$ and $\chi^{(3)}$ nonlinear processes.
    \textbf{c}. Inferred mmWave power in the generator CPS cavity as a function of the combined optical pump power in the bus waveguide. When the two pump powers are increased at a fixed ratio, the generated power exhibits the quadratic scaling expected for $\chi^{(2)}$ difference-frequency generation. The larger error bars at high pump powers reflect increased optical-cavity instability.
    \textbf{d}. Measured sideband power as a function of $\Omega/2\pi$. The top panel shows the \textit{ab initio} calculated CPS mode spectrum without (grey) and with loss (red), together with the sampled optical-mode separations (blue). The bottom panel shows the measured off-chip sideband power. Yellow markers show values obtained from the analytical model fitted to the measurements.
    \textbf{e}. Extracted drive-power-normalized on-chip photon-flux conversion efficiencies for generation (blue) and detection (red).
    \textbf{f}. Extracted single-photon electro-optic coupling rate $g_0/2\pi$ for this work (blue circles), compared with recent electro-optic implementations (crosses).
    }
    \label{Fig3}
\end{figure*}

We demonstrate the end-to-end operation of the system using the setup as shown in Fig.~\ref{Fig3}a, starting with $\Omega/2\pi=360$~GHz. Two resonant optical pumps deliver a combined power of approximately 5.6~mW to the generator bus waveguide, corresponding to an estimated circulating power of approximately 440~mW within the racetrack resonator. The generated 360~GHz field is routed to the detector module, where it modulates an optical probe to produce sidebands at $\omega_3\pm\Omega$ (Fig.~\ref{Fig3}b). The measured sideband has an off-chip power of approximately $-70$~dBm, from which we infer a mmWave-THz power of approximately 120~nW in the generator CPS cavity. With the probe power fixed and the two pump powers increased at a constant ratio, the generated power scales quadratically with their combined power (Fig.~\ref{Fig3}c), as expected for DFG. The power calibration and analytical model are detailed in the Supplementary Information.

We map the operating-frequency range of the signal core by pumping different pairs of racetrack modes and thereby tuning $\Omega$ discretely. The measured off-chip sideband powers are summarized in Fig.~\ref{Fig3}d, demonstrating end-to-end operation at sampled frequencies up to $\Omega/2\pi=2.56$~THz. Even at this highest frequency, the sideband remains approximately 2~dB above the local noise floor, with only 0.32~mW optical probe power at $\omega_3$ in the bus waveguide. The responses at 180, 360, and 540~GHz are approximately 10~dB stronger than those at neighboring sampled frequencies and coincide with the calculated CPS eigenfrequencies (Fig.~\ref{Fig3}d, upper panel), directly revealing the benefit of simultaneous triply resonant operation in the generator and detector modules. Away from the CPS resonances, such as at 432 and 468~GHz, the optical fields remain optically resonant and produce sidebands approximately 10~dB above the noise floor.

A model combining electro-optic conversion in both modules with the transfer function derived for the coupled CPS cavities reproduces the resonant peaks and overall response envelope (Fig.~\ref{Fig3}d and Supplementary Information). The overall response initially increases with frequency and reaches a maximum near 1~THz. For modes with comparable geometry and field overlap, this rise is consistent with the favorable scaling $g_0\propto\sqrt{\Omega}$, which originates from the increasing zero-point electric field. However, at higher frequencies, the growing CPS propagation loss (Fig.~\ref{Fig2}e) outweighs this scaling advantage and suppresses the response. Our measurements therefore expose a central design trade-off: the electro-optic interaction strengthens with frequency, while the on-chip high-frequency path becomes increasingly lossy. Reducing CPS loss is consequently the most direct route to further performance improvements.

We quantify the module-level performance using the drive-power-normalized photon-flux conversion efficiencies $\eta_{\mathrm{gen}}$ and $\eta_{\mathrm{det}}$ (Supplementary Information):
\begin{equation}
\begin{aligned}
\eta_\mathrm{gen}
= \frac{P_{\Omega,\mathrm{gen}}/\Omega}
{P_{\omega_1,\mathrm{bus}} \cdot P_{\omega_2,\mathrm{bus}}/\omega_2}
\qquad
\eta_\mathrm{det}
= \frac{P_\mathrm{SB,bus}/\omega_\mathrm{SB}}
{P_{\Omega,\mathrm{det}}/\Omega \cdot P_{\omega_3,\mathrm{bus}}}
\end{aligned}
\end{equation}

Here, $P_{\Omega,\mathrm{gen}}$ and $P_{\Omega,\mathrm{det}}$ are the mmWave-THz powers in the generator and detector CPS cavities, respectively, and all optical powers are referenced to the corresponding bus waveguides. For generation, the pump at $\omega_1$ serves as the drive, while the input at $\omega_2$ is converted to the mmWave-THz field. For detection, the probe at $\omega_3$ serves as the drive. The quantities $P_{\mathrm{SB,bus}}$ and $\omega_{\mathrm{SB}}$ are the power and angular frequency of the detected sideband, respectively.

The maximum efficiencies extracted reach $\eta_{\mathrm{gen}}=1.0\%~\mathrm{mW}^{-1}$ at $\Omega/2\pi=720$~GHz and $\eta_{\mathrm{det}}=0.5\%~\mathrm{mW}^{-1}$ at $\Omega/2\pi=864$~GHz (Fig.~\ref{Fig3}e). The resonant feature in $\eta_{\mathrm{gen}}$ arises from field buildup and interference within the generator CPS cavity. For $\eta_{\mathrm{det}}$, normalizing the sideband power $P_\mathrm{SB,bus}$ by the power inside the detector CPS cavity $P_{\Omega,\mathrm{det}}$ removes this buildup from the conversion metric. Above approximately 1.5~THz, the two efficiencies converge as propagation loss suppresses CPS resonant buildup. The system then approaches the nonresonant limit, in which the reciprocity of the underlying nonlinear optical interaction predicts equal generation and detection efficiencies.

Finally, we extract the single-photon electro-optic coupling rate $g_0$ (Supplementary Information) and compare it with recent electro-optic implementations~\cite{zhang2025ultrabroadband,zhang2025integrated,multani2026integrated,fan2018Superconducting,holzgrafe2020cavity,xu2021bidirectional,warner2025coherent}. Our measurements extend this benchmark into the terahertz regime, with $g_0/2\pi$ reaching the 10~kHz scale (Fig.~\ref{Fig3}f). At higher frequencies, the extracted $g_0/2\pi$ value exhibits an apparent roll-off as CPS propagation loss increases. In this high-loss regime, attenuation shortens the effective interaction length and reduces the THz energy stored within the electro-optic interaction region relative to the idealized cavity model. The model therefore overestimates the effective THz photon population that participates in the interaction and consequently underestimates $g_0/2\pi$. The apparent roll-off thus reflects a limitation of the extraction model rather than an intrinsic weakening of the electro-optic coupling.


\subsection*{4. Coherent control of mmWave-THz fields}

\begin{figure}[h!]
    \centering
    \includegraphics[width=1\textwidth]{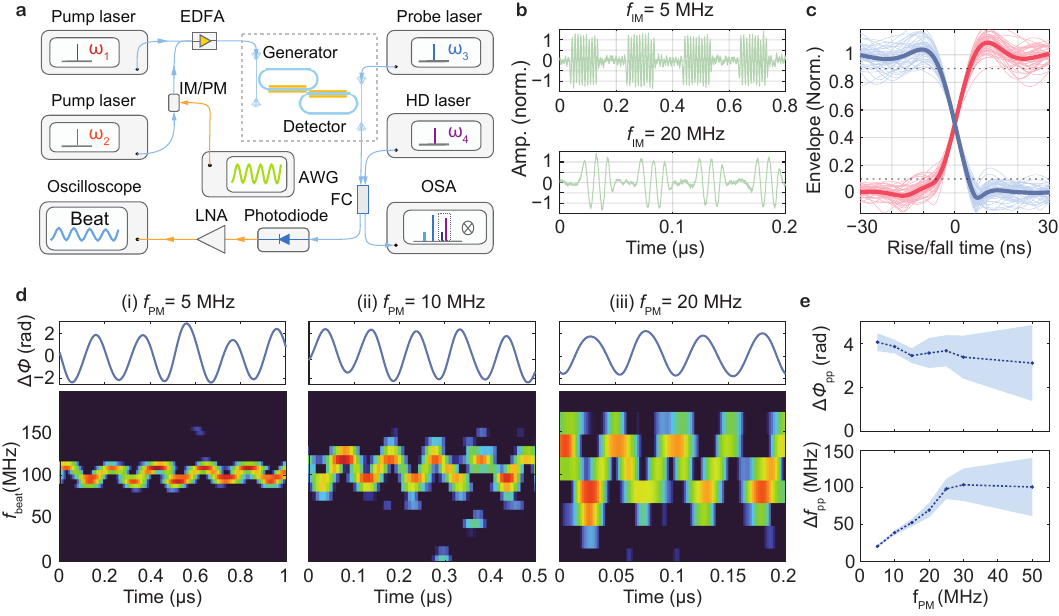}
    \renewcommand{\baselinestretch}{1} 
    \caption{\textbf{Programmable control of a mmWave-THz field at 360~GHz} 
    \textbf{a}. Schematic of the experimental setup for amplitude and phase control. An arbitrary waveform generator (AWG) drives an off-chip intensity modulator (IM) or phase modulator (PM), and the detector sideband is measured by heterodyne detection. EDFA, erbium-doped fiber amplifier; LNA, low-noise amplifier; HD, heterodyne detection; FC, fiber coupler. Optical and electrical paths are shown in blue and yellow, respectively.
    \textbf{b}. Temporal envelope of the heterodyne signal under square-wave intensity modulation at $f_{\mathrm{IM}}=5$ and $20~\mathrm{MHz}$ with a 50\% duty cycle. Traces are cropped to four modulation periods for clarity.
    \textbf{c}. Rising and falling transitions measured at $f_\mathrm{IM}=5$~MHz. The 10–90\% rise time is 10.3 ns, and the 90–10\% fall time is 9.2~ns. The light-blue and red traces show overlays of 50 rising and 49 falling transitions, respectively, extracted from the complete measurement record.
    \textbf{d}. Phase modulation under sinusoidal driving at $f_\mathrm{PM}=5$~MHz (i), 10~MHz (ii) and 20~MHz (iii). Top panels show the instantaneous phase of the heterodyne signal recovered using a Hilbert transform; bottom panels show the corresponding spectrograms, resolving instantaneous-frequency excursions about the heterodyne beat frequency.
    \textbf{e}. 
    Peak-to-peak phase excursion $\Delta\phi_{\mathrm{pp}}$ and instantaneous-frequency excursion $\Delta f_{\mathrm{pp}}$ as functions of $f_{\mathrm{PM}}$. For sinusoidal phase modulation, $\Delta f_{\mathrm{pp}}=\Delta\phi_{\mathrm{pp}}f_{\mathrm{PM}}$. The measured frequency excursion follows this relation up to approximately 25~MHz and approaches $\Delta f_{\mathrm{pp}}=100$~MHz. Deviations from ideal scaling are consistent with filtering by the finite optical-resonance linewidths.
    }
    \label{Fig4}
\end{figure}

We next exploit coherent electro-optic conversion to transfer the relative amplitude and phase of the optical pumps to the generated mmWave-THz field. We demonstrate this control at 360~GHz, which provides the highest signal-to-noise ratio among the measured operating frequencies (Fig.~\ref{Fig3}d). In the configuration shown in Fig.~\ref{Fig4}a, an arbitrary-waveform generator drives an off-chip electro-optic modulator that modulates one of the optical pumps. DFG transfers the applied modulation to the 360~GHz field, and subsequent upconversion in the detector maps it onto the optical sidebands. Heterodyne beating between one sideband and a reference laser produces an intermediate-frequency signal at $f_{\mathrm{beat}}\sim100$~MHz for time-resolved electrical analysis (Methods).

For amplitude control, we apply square-wave intensity modulation with a 50\% duty cycle. The recovered heterodyne envelope reproduces the programmed waveform at modulation frequencies $f_\mathrm{IM}=5$ and 20~MHz (Fig.~\ref{Fig4}b). The measured end-to-end response has a 10-90\% rise time of 10.3~ns and a 90-10\% fall time of 9.2~ns (Fig.~\ref{Fig4}c), demonstrating control of the 360~GHz field envelope on an approximately 10~ns timescale.

For phase control, we replace the intensity modulator with a phase modulator and apply sinusoidal modulation to one optical pump at $f_{\mathrm{PM}}=5$, 10, and 20~MHz. The recovered phase follows the applied modulation with a peak-to-peak excursion of $\Delta\phi_{\mathrm{pp}}\approx4$~rad (Fig.~\ref{Fig4}d, upper panels). The corresponding spectrograms reveal instantaneous beat-frequency excursions $\Delta f_{\mathrm{pp}}$ of approximately 20, 40, and 80~MHz, respectively (Fig.~\ref{Fig4}d, lower panels). Figure~\ref{Fig4}e summarizes both quantities over the full modulation-frequency sweep. The phase excursion remains nearly constant, with a modest high-frequency roll-off attributed to the frequency-dependent increase in the modulator $V_\pi$. Because instantaneous frequency is the time derivative of phase, the expected excursion is $\Delta f_{\mathrm{pp}}=\Delta\phi_{\mathrm{pp}}f_{\mathrm{PM}}$. The measurements follow this linear scaling up to $f_{\mathrm{PM}}=25$~MHz, where $\Delta f_{\mathrm{pp}}$ reaches approximately 100~MHz. At higher modulation frequencies, deviations from the ideal scaling indicate that the finite optical-cavity linewidth becomes limiting.

Together, these measurements demonstrate that the signal core functions not merely as a generator but as an optically programmable mmWave-THz waveform synthesizer. Such waveform agility is directly relevant to coherent communications and could support pulsed control of high-frequency quantum systems.


\subsection*{5. End-to-end cryogenic operation at 4~K}

\begin{figure}[h!]
    \centering
    \includegraphics[width=1\textwidth]{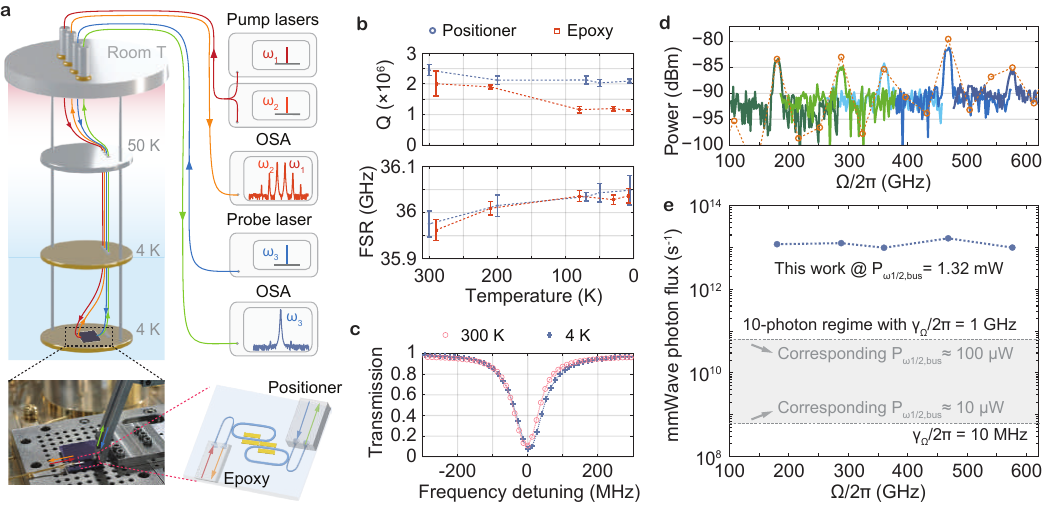}
    \renewcommand{\baselinestretch}{1} 
    \caption{\textbf{Cryogenic operation of the signal core.}
    \textbf{a}. Cryogenic measurement configuration and fiber interfaces to the photonic chip. A permanently bonded two-channel fiber array couples the optical pumps into the generator module, with a fiber-to-chip coupling efficiency of 22\%. A second fiber array, mounted on a cryogenic piezoelectric positioner, couples probe light into the detector module with an efficiency of 5\%, limited by the restricted angular adjustability of the positioner. The chip is mounted on the still plate of a dilution refrigerator and operated at approximately 4~K.
    \textbf{b}. Loaded optical quality factors (top) and free spectral ranges (bottom) of the epoxy-packaged generator racetrack (red squares) and positioner-aligned detector racetrack (blue circles) during cooling from 300 to approximately 4~K. At each temperature, the mean $Q$ is evaluated over approximately ten optical resonances, and the mean FSR over their adjacent mode spacings; the selected resonance pairs span the 100-600-GHz separations used in the experiments. Error bars denote the corresponding standard deviations.
    \textbf{c}. Representative transmission spectra of a detector-racetrack resonance near 1540~nm at room temperature (red circles) and 4~K (blue crosses). The resonator (positioner-aligned) remains close to critical coupling, with only a slight increase in linewidth upon cooling.
    \textbf{d}. Measured detector-output sideband signals at 180, 288, 360, 468, and 576~GHz. The yellow dashed curves represent analytical fits.
    \textbf{e}. Corresponding inferred mmWave-THz photon fluxes generated in the CPS cavity, reaching the order of $10^{13}$~photons~s$^{-1}$. The gray shaded region denotes the flux required to sustain a mean intracavity population $\bar{n}=10$ in a representative mmWave superconducting resonator with a loaded linewidth $\gamma_{\Omega}/2\pi$ between 10~MHz and 1~GHz. The estimated optical pump powers at the lower and upper boundaries are 10 and 100~$\mu\mathrm{W}$, respectively.
    }
    \label{Fig5}
\end{figure}

As a proof of concept, we mount a signal core on the still stage of a dilution refrigerator and demonstrate end-to-end cryogenic operation at approximately 4~K (Fig.~\ref{Fig5}a). Such demonstration is relevant for applications in, for instance, the dispersive readout of superconducting qubits, in which the qubit shifts the resonator frequency, and a weak coherent probe converts this shift into a measurable change in the phase or amplitude of the transmitted or reflected field~\cite{wallraff2005approaching,anferov2025millimeter}. Extending this approach to mmWave frequencies requires coherent tones to be generated, routed, and detected at cryogenic temperatures. The signal core performs these functions without external high-frequency electrical wiring and uses only fiber-based optical interfaces.

We implement two fiber-coupling schemes for the cryogenic signal chain. A two-channel fiber array is permanently bonded with epoxy to the generator side of the chip, whereas the detector is accessed through a fiber array mounted on a cryogenic micropositioner. Both interfaces retain alignment during assembly, vacuum pumping, and cooling, with insertion-loss variations below 1~dB (Supplementary Information). The detector racetrack maintains a loaded optical quality factor of approximately $2\times10^6$ during the cooling process, whereas the generator quality factor decreases to $1.3\times10^6$ before stabilizing (Fig.~\ref{Fig5}b,c). Because this degradation occurs only on the epoxy-bonded side, it may arise from surface contamination associated with epoxy outgassing, which could be addressed in the future by utilizing protective cladding. Importantly, the FSRs of the two racetracks remain matched throughout the cooling cycle, preserving the optical mode spacing required for simultaneous triply resonant operation from 300 to 4~K (Fig.~\ref{Fig5}b). 

At 4~K, a combined pump power of 2.6~mW in the generator bus waveguide (corresponding to approximately 77~mW of intracavity power) produces sidebands in the detector module at sampled frequencies ranging from $\Omega/2\pi=178$ to 576~GHz (Fig.~\ref{Fig5}d). At higher frequencies, the available amplifier-free pump power is insufficient to resolve the end-to-end response. 
The measured sideband powers correspond to inferred mmWave-THz powers on the order of nanowatts, or photon fluxes on the order of $10^{13}$~photons~s$^{-1}$, in the generator CPS cavity (Supplementary Information).

Dispersive readout commonly uses a low intracavity photon occupation to limit measurement backaction. Considering a resonator with a mean occupation $\bar{n}\sim10$ and energy-decay rates $\gamma_\Omega/2\pi=10$~MHz and 1~GHz, the corresponding photon-loss rates are $\bar{n}\gamma_\Omega\approx6.28\times10^8$ and $6.28\times10^{10}$~photons~s$^{-1}$, respectively. The demonstrated flux therefore exceeds these representative rates by approximately two to four orders of magnitude. Using the measured quadratic pump-power scaling, we estimate that fluxes in this range could be generated with combined bus-waveguide pump powers of approximately 10-100~\textmu W (Fig.~\ref{Fig5}e and Supplementary Information).

Although the present experiment does not incorporate a quantum device, it demonstrates the complete cryogenic signal chain required to optically generate, transfer, and recover coherent mmWave-THz signals. The estimated low-power operating regime further indicates compatibility with weak dispersive-readout tones and modest optical heat loads. The signal core therefore provides a fiber-interfaced photonic route to the control and readout of mmWave quantum hardware without an external high-frequency electrical signal chain.


\section*{Discussion}

The central advance of this work is a photonic-integrated architecture rather than an isolated converter, which enables end-to-end mmWave-THz signal chains with significantly reduced system complexity. By integrating electro-optic generation and detection modules with a resonant CPS link, the signal core treats mmWave-THz synthesis, routing, waveform control, and readout as co-designed functions on a single lithium tantalate chip. The same fiber-based interface supports cavity-selected operation from 100~GHz to 2.5~THz, nanosecond-scale waveform synthesis, and end-to-end operation at 4~K. Beyond these capabilities, the measurements establish a central design rule: although the electro-optic interaction strengthens with frequency, increasing conductor and radiation losses eventually outweigh this advantage. System performance is therefore governed by the balance among local conversion strength, inter-cavity coupling, and propagation loss rather than by any one component in isolation.

Two limitations define the next stage of development: spectral agility and high-frequency propagation loss. First, the triply resonant operating frequencies are constrained to integer multiples of the approximately 36-GHz optical free spectral range. Reconfigurable photonic molecules with controllable supermode splittings could introduce gigahertz-scale tuning around each operating point without sacrificing resonant enhancement~\cite{zhang2019electronically,zhang2026reconfigurable}. Second, as the CPS attenuation becomes stronger at higher frequencies, the CPS-CPS link crosses from resonant enhancement into the attenuation regime (Fig.~\ref{Fig2}c). Thicker metallization could reduce conductor loss, while ground-signal-ground geometries and low-refractive-index substrates could suppress radiative leakage and substrate-mode excitation~\cite{zhang2025monolithic}. At cryogenic temperatures, superconducting transmission lines provide an additional route to lower conductor loss and direct integration with superconducting circuits~\cite{anferov2020millimeter}, although this advantage is restricted to frequencies below the material-dependent pair-breaking threshold. Extending efficient operation further into the THz regime will therefore require joint optimization of conductor geometry, substrate properties, and electromagnetic mode confinement.

The demonstrated end-to-end operations at room and cryogenic temperatures point toward a more consequential use of the architecture: embedding a device under test directly within the CPS link. The generator and detector would then act as a local source and coherent analyzer, synthesizing, delivering, and recovering high-frequency fields without routing them through remote mmWave-THz instrumentation. In this paradigm, the experiment itself becomes part of the signal path, enabling in situ sensing and spectroscopy of integrated high-frequency devices. If added noise and stray optical absorption can be sufficiently suppressed, the same approach could interface with quantum hardware, including parametric amplifiers~\cite{qiu2023broadband,tan2024operation} and mmwave superconducting qubits~\cite{anferov2025millimeter}. More broadly, the concept of our signal core shifts high-frequency instrumentation from rack-level assemblies to fiber-addressed circuits co-located with the device of interest, offering a scalable route for next-generation systems in which wiring complexity, heat load, and footprint are the dominant constraints.


\section*{Methods}

\subsection*{Signal core fabrication}\label{methods:fab}

The mmWave/THz signal cores were fabricated on x-cut thin-film lithium tantalate (SIMIT, Shanghai), which consists of a 600~nm TFLT layer on a 4.7~\textmu m $\mathrm{SiO_x}$ buried oxide layer atop a high-resistivity silicon substrate. Photonic circuits were defined using an HSQ hard mask and electron-beam lithography, followed by development in 25\% TMAH. The patterns were then transferred into the TFLT layer by ion-beam etching (IBE) at 500~W with a 10$^\circ$ tilt angle for 350~nm etching depth, producing rib waveguides with an etch depth of 350~nm and a sidewall angle of approximately 75$^\circ$. After etching, lithium tantalate redeposition was removed using a 3:1 (by volume) mixture of 40\% KOH and 30\% $\mathrm{H_2O_2}$ at 85~$^\circ$C. 

To define the CPS and bowtie antenna structures, an MMA/PMMA bilayer was used as the lift-off resist and patterned by electron-beam lithography. A 5~nm Ti adhesion layer followed by 300~nm Au was deposited by electron-beam evaporation. The lift-off process was completed in acetone, after which the samples were cleaned with IPA and dried under $\mathrm{N_2}$ gas.

\subsection*{Optical and mmWave experiments} 

Optical transmission of the racetrack resonator is characterized using a tunable laser (Keysight N7776C) with an output power of 1~mW. Light is injected onto the chip using a fiber probe, coupled through a grating coupler and bus waveguide, and the transmitted signal is measured with a power meter (Keysight N7742C). The entire optical path uses polarization-maintaining fibers.

For optical pumping in the monolithic generation and detection experiments, two tunable lasers (Keysight N7776C) are combined using a fiber coupler (Thorlabs PW1550R5A2) and amplified by an erbium-doped fiber amplifier (Nuphoton EDFA-C0-PM-SMR-30-20-FCA) to a total output power of $\sim$100~mW. This results in an intracavity pump power of $\sim$250~mW for each tone. A third tunable laser (Keysight 8164B) is used as the probe. The output signal from the chip is collected with a fiber probe and analyzed using an optical spectrum analyzer (Anritsu MS9740B) with 0.03~nm resolution and 100~Hz video bandwidth.

RF experiments are performed using commercial RF multipliers (Virginia Diodes SGX WR9.0 and WR2.8) driven by an RF signal generator (Keysight E8267D), typically operated at 9--14~GHz and multiplied by factors of 9 or 27. This produces radiation powers of $\sim$100~mW and $\sim$4~mW in the 80--125~GHz and 260--380~GHz bands, respectively. The mmWave radiation is received by an on-chip bowtie antenna and coupled into the CPS, where it interacts with the optical field circulating in the racetrack resonator. This interaction induces phase modulation and generates optical sidebands, which are detected using the optical spectrum analyzer (Anritsu MS9740B).

\subsection*{Heterodyne experiments} 

The time-domain intensity-modulation experiments are conducted using an intensity modulator (Photline MXER-LN-10, typical $V_{\pi} \approx7$~V) driven by rectangular waveforms with 50\% duty cycle and $\sim$6~V peak-to-peak voltage from an RF generator (Agilent 33250A). This modulation is applied to one of the pump lasers (Keysight N7776C), resulting in intensity modulation of the generated mmWave-THz signal and, consequently, of the optical sideband. The sideband signal, with off-chip power on the order of a few nanowatts, is collected using a fiber probe and combined with a reference DBR laser (Aerodiode 1550LD) in a fiber coupler (Thorlabs PW1550R5A2), generating a beat note typically around 100~MHz. The reference-laser power is set to approximately 2~mW to maximize the beat signal without saturating the photodiode (Koheron PD100AC), which has a maximum gain of 3000~V/W. This results in a near on-off modulation with an amplitude on the order of 100~mV, recorded using a fast oscilloscope (Keysight EXR604A).

The phase-modulation experiments are conducted using a similar system, but with a phase modulator (Thorlabs LN53S-FC, typical $V_{\pi} \approx 3.5$~V at low modulation frequencies) replacing the intensity modulator. An arbitrary waveform generator (Zurich Instruments HDAWG) is employed as an RF drive, with a maximum peak-to-peak voltage of $V_\mathrm{m}=5~\mathrm{V}$, corresponding to a phase-modulation index $\beta=\pi \cdot \frac{V_\mathrm{m}}{V_\pi}\approx2.2$~rad and thus a peak-to-peak phase excursion $\Delta\phi=2\beta\approx 4.4$~rad. The optical sideband is combined with the reference laser and detected using a photoreceiver (Menlo Systems FPD510-FC-NIR; maximum conversion gain, $\sim10^5~\mathrm{V,W^{-1}}$), producing a heterodyne beat signal. The sideband power is approximately 1~nW, whereas the optical-carrier and reference-laser powers are each maintained at approximately 20~\textmu W. The total incident optical power therefore remains well below the saturation threshold of the photoreceiver. Under these conditions, the detected beat signal has an amplitude of several tens of millivolts. In these experiments, 10~\textmu s of time-domain data are recorded on the oscilloscope at a sampling rate of 16~GS/s. The data are band-pass filtered and then post-processed using a Hilbert transform to retrieve the phase information of the beat signal. The spectrogram is calculated using a Hann window and optimized FFT parameters.

\subsection*{Fiber array package} 

We permanently attach a fiber array to the grating-coupler pair used to pump and read out the generator racetrack resonator. A five-axis stage assembly (Thorlabs MBT401D and MBT501) is used to position the fiber-array unit (PM1550-XP fiber, Meisu Technology) and align it with the input and output grating couplers to maximize the transmitted optical power. Once the transmission is maximized, epoxy (EPO-TEK OG154-1, Epoxy Technology Europe AG) is applied to the chip at locations sufficiently far from the photonic devices. The fiber array is then realigned with the grating couplers to recover the maximum transmission. The epoxy is cured by illumination with 365-nm ultraviolet light (2PDS-E-UV-PEN, Epoxy Technology Europe AG) for 10~min. Finally, the chip is placed on a hot plate at $90~^{\circ}\mathrm{C}$ for 1~hour to complete the curing process.

\vspace{1cm}

\textbf{Data Availability}
The measurement data generated in this study will be deposited in the Zenodo database under https://doi.org/10.5281/zenodo.20443675

\medskip
\textbf{Code Availability}
The code used to plot the data within this paper will be deposited in the Zenodo database under https://doi.org/10.5281/zenodo.20443675

\section*{References}
\bibliographystyle{paper}
\bibliography{Reference}

\medskip
\textbf{Acknowledgments}
J.L. acknowledges funding through the Excellence Postdoctoral Fellowship Programme from the Quantum Science Center at EPFL and the Swiss National Science Foundation grant number 503051 (Swiss Quantum Call). T.Z. and G. J. acknowledge support from the Swiss National Science Foundation through grant number 515075 (SNSF-NSF lead agency project). A.G.P. acknowledges funding through the Excellence Postdoctoral Fellowship Programme from the Quantum Science Center at EPFL and the Swiss National Science Foundation grant number 515449 (SPARK). A.G. and Y.L. acknowledge funding from the European Union’s Horizon Europe research and innovation program under project MIRAQLS with grant agreement No 101070700. The chips were fabricated at the Center of MicroNanoTechnology (Cmi) at EPFL. 
H.L. acknowledges funding from the National Natural Science Foundation of China (62325509, 62505344, 62235019, 62531005, 62575299, 62275258, 62305364, 62435017, and T2550072). X.O. acknowledges the funding from the National Natural Science Foundation of China (No. 62293521)  and the National Key R\&D Program (2022YFA1404601) from the Ministry of Science and Technology of China.

\medskip
\textbf{Author contributions} J.L. and I.C.B.C. conceived the ideas; I.C.B.C. supervised the project; C.W. prepared the thin-film lithium tantalate samples under the supervision of X.O.; J.L. designed and fabricated the chip, built the experimental setups, performed the experiments, developed the theoretical models, and conducted the simulations; T.Z. and G.J assisted with the cryogenic measurements; A.G. and Y.L. assisted with the modulation experiments; A.G. and A.G.P. assisted with the theoretical calculations; A.T. assisted with simulations. J.L. analyzed the data in discussion with C.W., H.L., X.O. and I.C.B.C.; J.L. and I.C.B.C. wrote the manuscript with input from all authors.

\medskip
\textbf{Competing interests}
The authors declare no competing interests.

\medskip
\textbf{Corresponding authors}
The authors declare no competing interests. Correspondence to Jiawen Liu (jiawen.liu@epfl.ch), Xin Ou (ouxin@mail.sim.ac.cn) or Ileana-Cristina Benea-Chelmus (cristina.benea@epfl.ch).

\clearpage

\listoffigures

\end{document}


\renewcommand\refname{Supplementary References}

\maketitle
\newpage
\tableofcontents
\newpage

\section{Theoretical model for cavity-enhanced electro-optic generation and detection} \label{SI_theory}

\subsection{Sum-frequency generation (SFG) for mmWave-THz detection} 

We detect the mmWave-THz signal on chip by injecting an optical probe tone into a lithium tantalate device and generating sum-frequency sidebands through electro-optic conversion. Here, we derive the SFG process, closely following our previous work in Ref.~\cite{gaier2025wireless}.

We consider an optical probe field polarized along the x-axis, aligned with the extraordinary axis of the lithium tantalate crystal, and propagating along the y-axis. The optical probe field at frequency $\omega_{o3}$ can be written as
\begin{equation}
{E_\mathrm{o3}(x,y,z,t) = \frac{1}{2} A_\mathrm{o3}(y) g_\mathrm{opt}(x, z) e^{i(k_\mathrm{o3} y - \omega_\mathrm{o3} t)} + c.c.}
\end{equation}
where $A_{o3}(y)$ is the electric field amplitude, $k_\mathrm{o3}=n_\mathrm{eff} \frac{\omega_\mathrm{o3}}{c_0}$ is the propagation constant of the optical eigenmode with effective refractive index $n_\mathrm{eff}$, and $c_0$ is the speed of light in vacuum. Here, $g_\mathrm{opt}(x,z)$ is the 2D field profile of the optical waveguide eigenmode, normalized as $\iint |g_\mathrm{opt}|^2(x,z),dx,dz = S_\mathrm{opt}$, where $S_{opt}$ is the effective mode area.

For the mmWave-THz field to be detected, we write
\begin{equation} 
{E_\Omega(x,y,z,t) = \frac{1}{2} A^\mathrm{co}_{\Omega}(y) g_{\Omega}(x, z) e^{i(k_\Omega y - \Omega t)}e^{-\frac{\alpha}{2}y} +\frac{1}{2} A^\mathrm{coun}_{\Omega}(y) g_{\Omega}(x, z) e^{i(-k_\Omega y - \Omega t)}e^{\frac{\alpha}{2}y} + c.c.} 
\end{equation}
where $k_{\Omega}=n_{\Omega} \frac{\Omega}{c_0}$ is the wave vector of the mmWave-THz field with effective refractive index $n_{\Omega}$, $\alpha$ is the total mmWave-THz propagation loss, incorporating conductor, radiation and dielectric losses, and $g_{\Omega}$ is the mmWave-THz mode profile normalized as $\iint |g_{\Omega}|^2(x,z) dxdz = S_{\Omega}$. $A^{co}_{\Omega}$ and $A^\mathrm{coun}_{\Omega}$ are the amplitudes of the mmWave-THz fields that co- and counter-propagate with the optical probe inside the CPS cavity, respectively. These amplitudes are treated as slowly varying constants, while propagation loss is included explicitly through the exponential factors.

The wave equation governing the SFG process reads
\begin{equation}
\left(\nabla^2-\frac{1}{c_0^2}\frac{\partial^2}{\partial t^2}\right)E_\mathrm{SB}(x,y,z,t)
=
\frac{1}{\varepsilon_0 c_0^2}\frac{\partial^2}{\partial t^2}
\left(\mathcal{P}^{L}(x,y,z,t)+\mathcal{P}^{NL}(x,y,z,t)\right)
\end{equation}
with the linear polarization term
$\mathcal{P}^{L}(x,y,z,t)=\varepsilon_0 \chi^{(1)} \frac{1}{2} A_\mathrm{SB}(y)\,g_{\mathrm{opt}}(x,z)\,e^{i(k_\mathrm{SB}y-\omega_\mathrm{SB}t)}+\mathrm{c.c.}$, and the nonlinear term $\mathcal{P}^{NL}(x,y,z,t)=2\varepsilon_0 \chi^{(2)} E_\mathrm{o3}(x,y,z,t)\,E_{\Omega}(x,y,z,t)+\mathrm{c.c.}$

Solving this equation gives
\begin{equation}
{ \frac{\partial A_\mathrm{SB}(y)}{\partial y}  =
i\frac{\chi^{(2)} k_\mathrm{SB} }{2\varepsilon_r} A_\mathrm{o3}(y)\cdot \Gamma_\mathrm{eo} \cdot \left(A_{\Omega}^\mathrm{co}  e^{i\Delta k_\mathrm{co} y}e^{-\frac{\alpha}{2}y} + A_{\Omega}^\mathrm{coun} e^{i\Delta k_\mathrm{coun} y} e^{\frac{\alpha}{2}y} \right)}
\end{equation}
with the spatial overlap factor
${\Gamma_\mathrm{eo} = \frac{\iint_\mathrm{LN} g^2_\mathrm{opt} (x,z) \cdot g_{\Omega} (x,z), dx, dz}{\iint_{x, z} g^2_\mathrm{opt} (x,z), dx, dz}}$.

After integrating over the interaction length from $y=0$ to $L_\mathrm{cps}$ and assuming that the strong probe $A_{o3}$ remains approximately constant during the weak nonlinear interaction, the integral reduces to
\begin{equation} \label{Eq_SFG} 
A_\mathrm{SB}(L)=A_\mathrm{SB}(0)+i\,\frac{\chi^{(2)}k_\mathrm{SB}}{2\varepsilon_r}\,A_\mathrm{o3}(L)\Gamma_\mathrm{eo} \left[ A_{\Omega}^\mathrm{co}\,\frac{e^{\left(i\Delta k_\mathrm{co}-\frac{\alpha}{2}\right)L_\mathrm{cps}}-1}{i\Delta k_\mathrm{co}-\frac{\alpha}{2}} + A_{\Omega}^\mathrm{count}\,\frac{e^{\left(i\Delta k_\mathrm{count}+\frac{\alpha}{2}\right)L_\mathrm{cps}}-1}{i\Delta k_\mathrm{count}+\frac{\alpha}{2}} \right] 
\end{equation}
which can be further simplified to 
\begin{equation} \label{Eq_SFG_simple} 
A_\mathrm{SB}=i\,\frac{\chi^{(2)}k_\mathrm{SB}}{\varepsilon_r}\,A_\mathrm{o3}\Gamma_\mathrm{eo} A_{\Omega}^\mathrm{co}\,\frac{1-e^{\left(-\frac{\alpha}{2}\right)L_\mathrm{cps}}}{\alpha} 
\end{equation}
assuming perfect phase matching and negligible effect from the counter-propagation term.

\newpage
\subsection{Difference-frequency generation (DFG) for mmWave-THz generation} 

We now extend the model to mmWave-THz signal generation using difference-frequency generation (DFG).
The optical pump fields can be written as
\begin{equation}
    {E_\mathrm{o1,2}(x,y,z,t) = \frac{1}{2} A_\mathrm{o1,2}(y) g_\mathrm{opt}(x, z) e^{i(k_\mathrm{o1,2} y - \omega_\mathrm{o1,2} t)} + c.c.}
\end{equation}
where the subscripts $\mathrm{o1,2}$ represent the two optical pump fields, respectively, $A_\mathrm{o1,2}$ is the electric field amplitude, $k_\mathrm{o1,2}=n_\mathrm{eff} \frac{\omega_{1,2}}{c_0}$ is the propagation constant of the optical eigenmode.

For the generated mmWave-THz field, we can write:
\begin{equation}
    {E_\Omega(x,y,z,t) = \frac{1}{2} A_{\Omega}(y) g_\Omega(x, z) e^{i(k_\Omega y - \Omega t)} + c.c.}
\end{equation}
where $A_{\Omega}$ is the amplitude of the generated mmWave-THz field, and $k_\Omega=n_{\Omega} \frac{\Omega}{c_0}$ is the propagation constant of the mmWave-THz eigenmode with effective refractive index $n_{\Omega}$.

For the DFG process, we write
\begin{equation}
    \left(\nabla^2 - \frac{1}{c_0^2}\frac{\partial^2}{\partial t^2}\right) E_{\Omega}(x,y,z,t)
    = \frac{1}{\varepsilon_0 c_0^2}\frac{\partial^2}{\partial t^2}
    \left(\mathcal{P}^{L}(x,y,z,t) + \mathcal{P}^{NL}(x,y,z,t)\right)
\end{equation}
where ${\mathcal{P}^L(x,y,z,t) = \varepsilon_0 \chi^{(1)} \frac{1}{2} A_{\Omega}(y)g_\Omega(x, z) e^{i(k_\Omega y - \Omega t)}}+c.c.$ is the linear part of the polarization, with ${\chi^{(1)}}$ being the linear susceptibility, and ${\mathcal{P}^{NL}(x,y,z,t)  = 2 \varepsilon_0 \chi^{(2)}_{333}(x,y,z,t)\cdot E_{o1}^*(x,y,z,t)E_{o2}(x,y,z,t)}+c.c.$ is the nonlinear part of the polarization, with ${\chi^{(2)}_{333}}$ being the second-order nonlinear susceptibility.

Solving the wave equation, we find
\begin{equation}
    { \frac{\partial A_{\Omega}(y)}{\partial y} (ik_\Omega) e^{i(k_\Omega y)} =
    -\frac{\chi^{(2)} \Omega^2 }{2c_0^2} A_\mathrm{o1}(y)A_\mathrm{o2}(y) \Gamma_\mathrm{oe} \cdot e^{i\Delta k_\mathrm{DFG}\cdot y}}
\end{equation}
with phase-matching term $\Delta k_{\mathrm{DFG}} = k_\mathrm{o2} - k_\mathrm{o1} - k_{\Omega}$.

If we include the THz mode loss term $e^{-\frac{\alpha}{2} y}$, then the propagation equation (with no drive) is $\frac{\partial A_{\Omega}(y)}{\partial y} = -\frac{\alpha}{2} A_{\Omega}(y)$. Adding this loss term to the driven equation gives
\begin{equation}
    \frac{\partial A_{\Omega}(y)}{\partial y} + \frac{\alpha}{2} A_{\Omega}(y)
    = i\,\frac{\chi^{(2)} \Omega}{2 n c_0}\,A_\mathrm{o1}(y)\,A_\mathrm{o2}(y)\,\Gamma_\mathrm{oe}\,e^{i(\Delta k_{\mathrm{DFG}})y}
\end{equation}
After integrating over the interaction length from $y=0$ to $L_\mathrm{cps}$ and assuming that the strong pumps $A_\mathrm{o1,2}$ remain approximately constant during the weak nonlinear interaction, we eventually obtain the amplitude of the generated mmWave--THz signal:
\begin{equation} 
    A_{\Omega}(y=L_\mathrm{cps})
    =
    e^{-\frac{\alpha}{2}L}
    \left[
    A_{\Omega}(0)
    +
    i\,\frac{\chi^{(2)}\Omega}{2nc_0}\,\Gamma_\mathrm{oe}\,A_\mathrm{o1}(L)\,A_\mathrm{o2}(L)\,
    \frac{e^{\left(i\Delta k_{\mathrm{DFG}}+\frac{\alpha}{2}\right)L_\mathrm{cps}}-1}
    {i\Delta k_{\mathrm{DFG}}+\frac{\alpha}{2}}
    \right]
\end{equation}
with $A_{\Omega}(0)=0$ in the absence of an initial mmWave-THz signal on chip. If we assume near-perfect phase matching and strong non-depleting optical pump, eventually one can write:
\begin{equation}\label{Eq_DFG}
A_{\Omega}(y=L_\mathrm{cps})
= i\,\frac{\chi^{(2)}\Omega}{2 n_\Omega c_0}\,
\Gamma_{oe} A_\mathrm{o1} A_\mathrm{o2} \cdot \frac{1 - e^{-\frac{\alpha}{2}L_\mathrm{cps}}}{\alpha/2}.
\end{equation}

\newpage
\subsection{CPS cavity enhancement for mmWave-THz generation and detection}

\begin{figure}[h]
    \centering
    \includegraphics[width=1\textwidth]{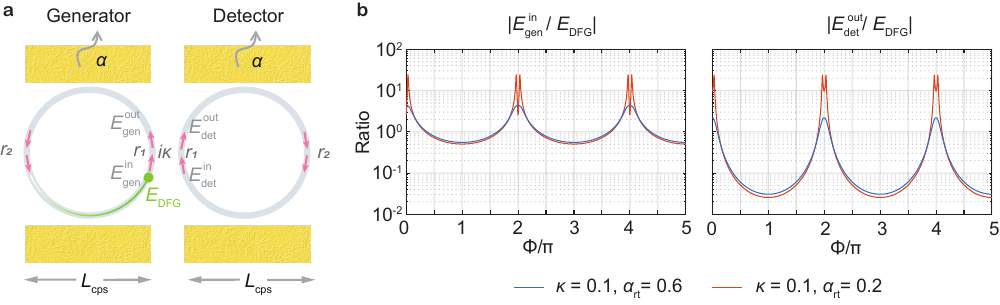}
    \caption{
    \textbf{Cavity-enhanced generation and transfer of mmWave-THz fields.}
    \textbf{a}. Schematic of the coupled-CPS-cavity model used to describe generation, on-chip transfer, and detection of the mmWave-THz field. 
    \textbf{b}. Calculated field ratios $E_\mathrm{gen}^\mathrm{in}/E_{\mathrm{DFG}}$ and $E_\mathrm{det}^\mathrm{out}/E_{\mathrm{DFG}}$ as functions of the propagation phase $\phi$, respectively. Both ratios show resonance-dependent enhancement due to the CPS cavities: constructive interference enhances the mmWave-THz field on resonance, whereas destructive interference suppresses the field off resonance. In the low-loss case, $E_\mathrm{gen}^\mathrm{in}/E_{\mathrm{DFG}}$ and $E_\mathrm{det}^\mathrm{out}/E_{\mathrm{DFG}}$ are comparable, indicating efficient inter-cavity field transfer. In contrast, in the high-loss case, propagation loss suppresses the field reaching the detector cavity, leading to a substantial difference between the two ratios.
    }
    \label{FigS_A1}
\end{figure}

As shown in Supplementary Fig.~\ref{FigS_A1}a, the mmWave-THz field is generated through the DFG process over the interaction length $L_\mathrm{cps}$. In the coupled-cavity model, we replace this distributed generation process by an equivalent lumped intracavity source with field amplitude $E_\mathrm{DFG}=A_{\Omega}(L_\mathrm{cps})$, located at the right termination of the generation CPS cavity. This effective source represents the accumulated single-pass mmWave-THz field generated by DFG and allows us to calculate the cavity enhancement arising from its circulation inside the coupled CPS resonators.

The two CPS cavities are capacitively coupled with inter-cavity field-coupling coefficient $i\kappa$, where the factor $i$ accounts for the extra phase shift introduced by the coupling interface. The fields defined in Supplementary Fig.~\ref{FigS_A1} satisfy
\begin{align} 
E_\mathrm{gen}^\mathrm{out} &= i\kappa E_\mathrm{det}^\mathrm{in} + r_1 E_\mathrm{gen}^\mathrm{in}, \\
E_\mathrm{det}^\mathrm{in} &= r_2 E_\mathrm{det}^\mathrm{out} e^{-\frac{\alpha}{2}\cdot 2L_\mathrm{cps}} e^{i\phi}, \\
E_\mathrm{det}^\mathrm{out} &= i\kappa E_\mathrm{gen}^\mathrm{in} + r_1 E_\mathrm{det}^\mathrm{in}, \\
E_\mathrm{gen}^\mathrm{in} &= r_2 E_\mathrm{gen}^\mathrm{out} e^{-\frac{\alpha}{2}\cdot 2L_\mathrm{cps}} e^{i\phi} + E_\mathrm{DFG},
\end{align}
where $r_1$ and $r_2$ are field-reflection coefficients, $\kappa$ is the inter-cavity field-coupling coefficient, $\phi$ is the round-trip propagation phase, and $\alpha$ is the CPS power-loss coefficient.

Solving these coupled equations gives the enhanced mmWave-THz field in the generation CPS cavity:
\begin{equation} \label{Eq_E1Ein}
E_\mathrm{gen}^\mathrm{in} =
\frac{
1 - r_1 r_2 e^{-\frac{\alpha}{2}\cdot 2L_\mathrm{cps}} e^{i\phi}
}{
1 - 2r_1 r_2 e^{-\frac{\alpha}{2}\cdot 2L_\mathrm{cps}} e^{i\phi}
+ r_2^2 e^{-\alpha \cdot 2L_\mathrm{cps}} e^{2i\phi}
}
E_{\mathrm{DFG}}
\end{equation}
as well as the field transferred to the detector CPS cavity
\begin{equation}  \label{Eq_E4Ein}
E_\mathrm{det}^\mathrm{out} =
\frac{
i\kappa
}{
1 - 2r_1 r_2 e^{-\frac{\alpha}{2}\cdot 2L_\mathrm{cps}} e^{i\phi}
+ r_2^2 e^{-\alpha \cdot 2L_\mathrm{cps}} e^{2i\phi}
}
E_{\mathrm{DFG}}
\end{equation}

We plot the field ratios $E_\mathrm{gen}^\mathrm{in}/E_\mathrm{DFG}$ and $E_\mathrm{det}^\mathrm{out}/E_\mathrm{DFG}$ as functions of the propagation phase $\phi$, using Eqs.~(\ref{Eq_E1Ein}) and~(\ref{Eq_E4Ein}), respectively. Supplementary Fig.~\ref{FigS_A1}b shows the calculated field transfer functions for different values of field-coupling coefficient $\kappa$ and round-trip field-attenuation factor $\alpha_\mathrm{rt}$. The blue curve corresponds to the signal-core case at 100~GHz, whereas the orange curve represents a low-loss case. On resonance, constructive interference enhances the mmWave-THz field generated through the DFG process, while off resonance the field is suppressed by destructive interference. Comparing blue and orange curves, we clearly see that, in the low-loss case, $E_\mathrm{gen}^\mathrm{in}/E_\mathrm{DFG}$ and $E_\mathrm{det}^\mathrm{out}/E_\mathrm{DFG}$ are comparable, indicating efficient transfer between the two CPS cavities, while in the high-loss case, the two ratios differ substantially because propagation loss suppresses the field transferred to the detector cavity.

These results showcase that CPS cavity resonances can enhance the generated and transferred mmWave-THz fields, provided that the inter-cavity coupling is sufficiently strong and the propagation loss is low.

\newpage
\subsection{Combined model for cavity-enhanced mmWave-THz signal generation and detection} 

We now combine all the previous derivations, from mmWave-THz signal generation via DFG in the generation module, through field transfer efficiency between coupled CPS cavities, to signal detection via SFG in the detection module. 

From equation~(\ref{Eq_DFG}) and (\ref{Eq_E1Ein}), after considering cavity effects, the generated mmWave-THz field amplitude in the generator module can be written as
\begin{equation}\label{Eq_gen}
A_{\Omega,{gen}}
= i \frac{\chi^{(2)} \Omega}{2 n_\Omega c_0}
\Gamma_{oe} A_{o1} A_{o2}
\frac{1 - e^{-\frac{\alpha}{2}\cdot L_\mathrm{cps}}}{\alpha/2}
\frac{
1 - r_1 r_2 e^{-\frac{\alpha}{2}\cdot 2L_\mathrm{cps}} e^{i\phi}
}{
1 - 2 r_1 r_2 e^{-\frac{\alpha}{2}\cdot 2L_\mathrm{cps}} e^{i\phi}
+ r_2^2 e^{-\alpha \cdot 2L_\mathrm{cps}} e^{2i\phi}
}
\end{equation}
while using equation~(\ref{Eq_DFG}) and (\ref{Eq_E4Ein}) the on-chip field transferred to the detector module can be written as:
\begin{equation}\label{Eq_det}
A_\mathrm{\Omega,{det}}
= i \frac{\chi^{(2)} \Omega}{2 n_\Omega c_0}
\Gamma_\mathrm{oe} A_{o1} A_{o2}
\frac{1 - e^{-\frac{\alpha}{2}\cdot L_\mathrm{cps}}}{\alpha/2}
\frac{
i\kappa
}{
1 - 2r_1 r_2 e^{-\frac{\alpha}{2}\cdot 2L_\mathrm{cps}} e^{i\phi}
+ r_2^2 e^{-\alpha \cdot 2L_\mathrm{cps}} e^{2i\phi}
}
\end{equation}

By combining equations~(\ref{Eq_SFG}) and (\ref{Eq_det}), we obtain the detected sideband signal amplitude in the detector module:
\begin{equation}\label{Eq_A_SB}
A_\mathrm{SB}
=
\frac{\left(\chi^{(2)}\right)^2 k_\mathrm{SB}\Omega}
{n_\Omega c_0 \varepsilon_r}
A_{o1}A_{o2}A_{o3}\Gamma_\mathrm{eo}\Gamma_\mathrm{oe}
\frac{
2e^{-\alpha L_\mathrm{cps}/2} - 1 - e^{-\alpha L_\mathrm{cps}}
}{\alpha^2}
\frac{
ik
}{
1 - 2 r_1 r_2 e^{-\frac{\alpha}{2}\cdot 2L_\mathrm{cps}} e^{i\phi}
+ r_2^2 e^{-\alpha \cdot 2L_\mathrm{cps}} e^{2i\phi}
}
\end{equation}
where we neglect the counter-propagation term in equation~(\ref{Eq_SFG}) and assume nearly perfect phase matching for the co-propagation term, which is reasonable for a very short CPS length of $750~\mu m$, while the coherent length is on the millimeter scale in the mmWave-THz range.

Then, the on-chip power of the mmWave-THz signal can be calculated as
\begin{equation}\label{Eq_THzPower}
    P_{\Omega,\mathrm{gen/det}}=\frac{1}{2}\varepsilon_{0} c_{0} n_{\Omega}
    \left|A_{\Omega,\mathrm{gen/det}}\right|^{2} S_{\Omega},
\end{equation}
where the subscript $\mathrm{gen/det}$ denotes the mmWave-THz field in the generator or detector module, respectively, and 
$S_{\Omega}=\iint \left|g_{\Omega}(x,z)\right|^{2}dx\,dz$ represents the effective cross-sectional mode area of the mmWave-THz mode inside the CPS.

Similarly, the sideband signal power can be calculated as
\begin{equation}\label{Eq_P_SB}
    P_\mathrm{SB}=\frac{1}{2}\varepsilon_{0} c_{0} n \left|A_\mathrm{SB}\right|^{2} S_\mathrm{opt},
\end{equation}
with $S_\mathrm{opt}=\iint \left|g_\mathrm{opt}(x,z)\right|^{2}dxdz$ being the effective cross-sectional mode area of the optical mode inside the LiTaO$_3$ waveguide.

Note that all optical fields and powers discussed in this section refer to intracavity values. In the next section, we characterize the optical cavity response of the racetrack resonator and extract both the field-enhancement factors and the grating-coupler coupling efficiency, which are required for accurate calibration of the on-chip powers.

\newpage
\subsection{Optical transmission characterization and resonator-enhancement estimation}

\begin{figure}[h!]
\centering
\includegraphics[width=\textwidth]{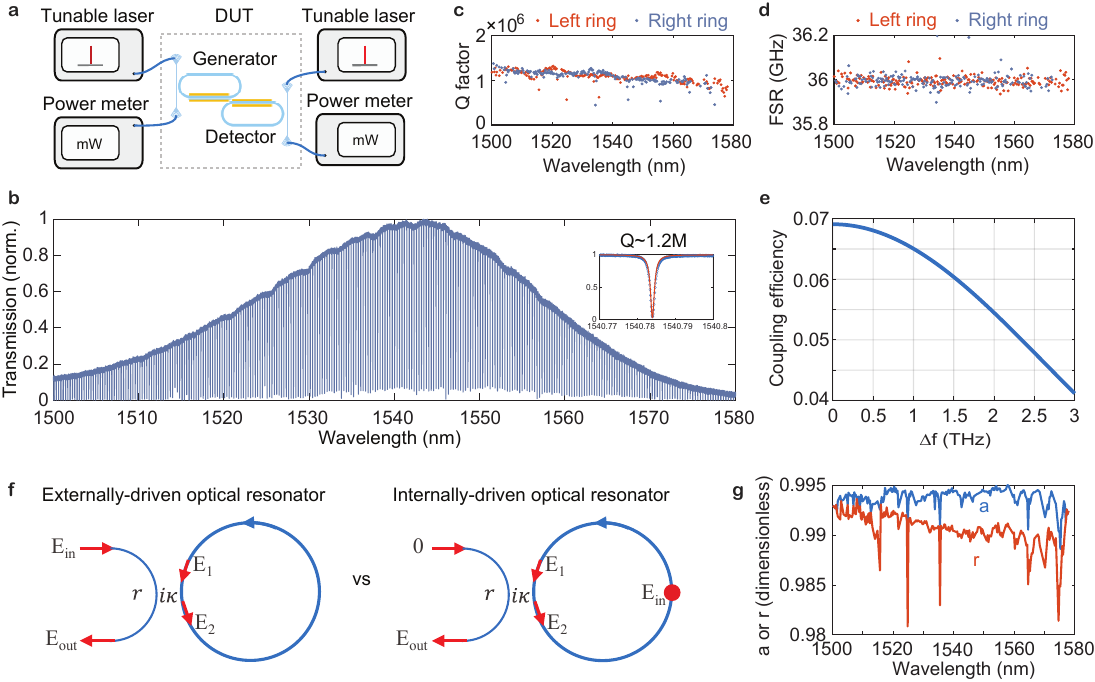}
\caption{\textbf{Optical characterization and resonator-enhancement estimation.}
\textbf{a}. Experimental setup used to characterize the racetrack-resonator transmission spectra and grating-coupler coupling efficiency. DUT, device under test. Blue lines indicate optical fiber interconnections.
\textbf{b}. Representative normalized transmission spectrum of a racetrack resonator. The near-Gaussian envelope is set by the wavelength-dependent grating-coupler response, while the resonance notches approaching zero transmission indicate near-critical coupling to the racetrack resonator. Inset: representative Lorentzian fit to one resonance mode, yielding a loaded optical $Q$ factor of $1.2\times10^6$.
\textbf{c}. Extracted loaded $Q$ factors for the racetrack resonators in the generator (orange) and detector (blue) modules.
\textbf{d}. Extracted free spectral ranges of the two racetrack resonators. Blue: $\mathrm{FSR=35.995 \pm 0.025~GHz}$. Red: $\mathrm{FSR=35.994 \pm 0.022~GHz}$.
\textbf{e}. Extracted grating-coupler coupling efficiency as a function of optical frequency offset. This offset corresponds to the pump-laser separation $\omega_1-\omega_2$ in the generation module and to the carrier-sideband separation between $\omega_3$ and $\omega_3\pm\Omega$ in the detection module.
\textbf{f}. Coupled-field models used to calculate the resonator field-enhancement factors. Left: generator module, where the optical pump is coupled into the racetrack resonator from the bus waveguide. Right: detection module, where the optical sideband is generated inside the resonator through electro-optic interaction and treated as an intracavity optical source.
\textbf{g}. Extracted round-trip field transmission coefficient $a$ and field self-coupling coefficient $r$ across the measured wavelength range. The extracted values, $a=0.994\pm0.001$ and $r=0.992\pm0.002$, are used to calculate the resonator field-enhancement factors in the generator and detector modules.
}
\label{FigS_A5}
\end{figure}

Supplementary Fig.~\ref{FigS_A5}a illustrates the experimental setup used to characterize the racetrack resonances and grating-coupler coupling efficiency. A representative normalized transmission spectrum is shown in Supplementary Fig.~\ref{FigS_A5}b. The spectrum exhibits a near-Gaussian envelope set by the wavelength-dependent grating-coupler response, together with resonance notches that approach zero transmission, indicating near-critical coupling to the racetrack resonator across the measured band. The inset shows a representative Lorentzian fit to one resonance mode, yielding a loaded optical $Q$ factor of $1.2\times10^6$. 

We then fit all resonance modes in Supplementary Fig.~\ref{FigS_A5}b with Lorentzian functions to extract the resonance frequencies $f_0$ and linewidths $\Delta f$. From these values, we calculate the loaded $Q$ factors, $Q=f_0/\Delta f$, and the free spectral ranges, as summarized in Supplementary Fig.~\ref{FigS_A5}c,d. The two racetrack resonators in generator and detector modules exhibit loaded $Q$ factors near $1.2\times10^6$ and nearly identical FSRs of approximately 36~GHz. Across an 80~nm optical spectral range, the FSR standard deviation remains below 25~MHz, corresponding to less than 0.07\% variation and well below the optical linewidth of approximately 150~MHz. This spectral uniformity facilitates simultaneous satisfaction of the triply resonant condition in both modules, as required for multifunctional operation on a scalable integrated photonic platform.

The same transmission spectrum is also used to extract the grating-coupler coupling efficiency, which is critical for on-chip power calibration. We fit the transmission envelope with a Gaussian function and obtain a peak transmitted power of approximately $3~\mu\mathrm{W}$ for an input pump power of $700~\mu\mathrm{W}$. Assuming identical input and output grating couplers, this corresponds to a peak per-coupler efficiency of approximately 7\%, which decreases as the optical frequency moves away from the grating-coupler center frequency. Supplementary Fig.~\ref{FigS_A5}e summarizes the extracted coupling efficiency as a function of optical frequency offset. For the generation module, this offset corresponds to the separation $\omega_1-\omega_2$ between the two pump lasers. For the detection module, it corresponds to the separation between the optical carrier $\omega_3$ and the sidebands $\omega_3\pm\Omega$.

We next consider the field-enhancement effect of the optical resonator in the generator module, where we pump the ring resonator with bus waveguide. As shown schematically in the left panel of Supplementary Fig.~\ref{FigS_A5}f, the steady-state field relations can be written as
\begin{align}
&E_2 = i k E_{\mathrm{in}} + rE_1, \\
&E_1 = a e^{i\phi} E_2, \\
&E_{\mathrm{out}} = i k E_1 + rE_{\mathrm{in}} .
\end{align}
Here, $r$ and $k$ are the field self-coupling and cross-coupling coefficients of the coupler, respectively, and $a = e^{-\frac{\alpha}{2}L}$ is the round-trip field transmission coefficient. The parameter $\alpha$ denotes the optical power-attenuation coefficient, $L$ is the total racetrack circumference, and $\phi$ is the round-trip propagation phase.

Solving these coupled-field relations gives the intracavity field enhancement factor
\begin{equation}\label{Eq_FE_gen}
\frac{E_2}{E_{\mathrm{in}}} = \frac{i k}{1-a r e^{i\phi}} .
\end{equation}
and the corresponding through-port field transmission
\begin{equation}
\frac{E_{\mathrm{out}}}{E_{\mathrm{in}}}
= r - \frac{k^2 a e^{i\phi}}{1-a r e^{i\phi}} .
\end{equation}
For a lossless coupler satisfying $r^2+k^2\approx1$, this expression reduces to
\begin{equation}\label{Eq_transT}
\frac{E_{\mathrm{out}}}{E_{\mathrm{in}}}
= \frac{r-ae^{i\phi}}{1-a r e^{i\phi}} .
\end{equation}

In the detection module, by contrast, the optical sideband is generated inside the ring through electro-optic interaction rather than being coupled from the bus waveguide. It can therefore be treated as an intracavity optical source, as shown schematically in the right panel of Supplementary Fig.~\ref{FigS_A5}f. The corresponding steady-state field relations are
\begin{align}
    &E_2 = rE_1, \\
    &E_1 = a e^{i\phi} E_2 + E_\mathrm{in}, \\
    &E_\mathrm{out} = i t E_1 .
\end{align}
where $E_\mathrm{in}$ corresponds to $A_\mathrm{SB}$ in Eq.~(\ref{Eq_SFG_simple}).

Solving these coupled-field relations gives the intracavity field enhancement factor
\begin{equation}\label{Eq_FE_det}
    \frac{E_2}{E_\mathrm{in}} = \frac{r}{1-a r e^{i\phi}}
\end{equation}
and the corresponding through-port field transmission
\begin{equation}
    \frac{E_{\mathrm{out}}}{E_{\mathrm{in}}} = \frac{i t}{1-a r e^{i\phi}}
\end{equation}

In addition, according to the ring resonator theory, the quality factor Q can be calculated as 
\begin{equation}\label{Eq_Q}
    Q = \frac{\sqrt{a r}}{1-a r} \cdot \frac{\pi \lambda}{FSR}
\end{equation}
Using Eqs.~(\ref{Eq_transT}) and~(\ref{Eq_Q}), together with the measured resonance extinction in Supplementary Fig.~\ref{FigS_A5}b and the extracted loaded $Q$ factors in Supplementary Fig.~\ref{FigS_A5}c, we calculate the corresponding round-trip field transmission coefficient $a$ and field reflection coefficient $r$. As shown in Supplementary Fig.~\ref{FigS_A5}g, across the measured wavelength range, we obtain $a=0.994\pm0.001$ and $r=0.992\pm0.002$. The gradual decrease of $r$ toward longer wavelengths indicates stronger evanescent coupling to the racetrack resonator on the long-wavelength side.

Using the extracted values of $a$ and $r$, we calculate the field-enhancement factors from Eqs.~(\ref{Eq_FE_gen}) and~(\ref{Eq_FE_det}). We obtain a 9-fold field enhancement, corresponding to an 81-fold power enhancement, in the generator module, and a 71-fold field enhancement, corresponding to a 5000-fold power enhancement, in the detection module.

\newpage

\subsection{Electro-optic and Optical-to-electrical mode overlap}

\begin{figure}[h!]
    \centering
    \includegraphics[width=1\textwidth]{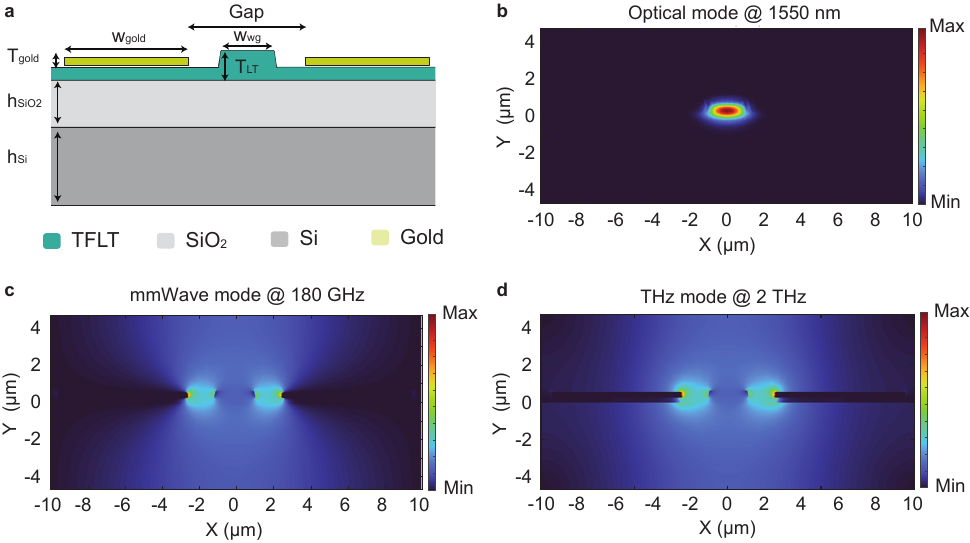}
    \caption{\textbf{Electro-optic mode-overlap simulation.}
    \textbf{a}. Simulated cross-sectional geometry of the TFLN waveguide and CPS electrodes used to calculate the optical and mmWave-THz mode profiles.
    \textbf{b-d}. Simulated $x$-component electric-field profiles for the optical mode at 1550~nm, the mmWave mode at 180~GHz, and the terahertz mode at 2~THz, respectively. The mmWave-THz fields remain confined within the electrode gap over this frequency range. From the normalized mode profiles, we extract $S_\mathrm{opt}\approx1~\mu\mathrm{m}^2$, $S_\Omega\approx31~\mu\mathrm{m}^2$ at 180~GHz, and $S_\Omega\approx29~\mu\mathrm{m}^2$ at 2~THz, yielding overlap factors of $\Gamma_\mathrm{eo}\approx0.6$ and $\Gamma_\mathrm{oe}\approx0.02$ across the mmWave-THz range.
    }
    \label{FigS_A6}
\end{figure}

\begin{table}[h]
\centering
\caption{\textbf{Device dimensions.}}
\label{Table_dim}
\begin{tabular}{lll}
\hline
Parameter & Symbol & Value \\
\hline
Lithium tantalate thickness & \(h_\mathrm{LT}\) & \(600~\mathrm{nm}\) \\
Buried oxide thickness & \(h_{\mathrm{SiO}_2}\) & \(4.7~\mu\mathrm{m}\) \\
Silicon substrate thickness & \(h_\mathrm{Si}\) & 500~\(\mu\mathrm{m}\) \\
Waveguide width & \(w_\mathrm{wg}\) & \(1.5~\mu\mathrm{m}\) \\
Gold thickness & \(T_\mathrm{gold}\) & \(150~\mathrm{nm}\) \\
Stripline gap & \(Gap\) & \(5~\mu\mathrm{m}\) \\
Gold strip width & \(w_\mathrm{gold}\) & 7~\(\mu\mathrm{m}\) \\
\hline
\end{tabular}
\end{table}

To calculate the on-chip mmWave-THz and optical powers using the theory derived above, we also need the electro-optic mode-overlap factors: $\Gamma_\mathrm{eo}$ for mmWave-THz-to-optical up-conversion and $\Gamma_\mathrm{oe}$ for optical-to-mmWave-THz down-conversion. These factors quantify the spatial overlap between the optical mode and the mmWave-THz mode. We therefore calculate the overlap between the simulated optical electric field, denoted as $E_{\mathrm{opt}}(x,z)$, and the simulated  mmWave-THz electric field, denoted as $E_{\Omega}(x,z)$.
The effective mode areas of the microwave and optical modes are defined as
\begin{equation}\label{Eq_S_Om}
S_{\Omega}
=
\frac{
\left(
\displaystyle \iint_{xz} |E_{\Omega}(x,z)|^2 \, dx\,dz
\right)^2
}{
\displaystyle \iint_{xz} |E_{\Omega}(x,z)|^4 \, dx\,dz
},
\end{equation}
and
\begin{equation}\label{Eq_S_opt}
S_{\mathrm{opt}}
=
\frac{
\left(
\displaystyle \iint_{xz} |E_{\mathrm{opt}}(x,z)|^2 \, dx\,dz
\right)^2
}{
\displaystyle \iint_{xz} |E_{\mathrm{opt}}(x,z)|^4 \, dx\,dz
}.
\end{equation}

Using these effective mode areas, the root-mean-square field amplitudes are calculated as
\begin{equation}
A_\Omega
=
\sqrt{
\frac{
\displaystyle \iint_{xz} |E_{\Omega}(x,z)|^2 \, dx\,dz
}{
S_{\Omega}
}
},
\end{equation}
for mmWave-THz modes, and
\begin{equation}
A_{\mathrm{opt}}
=
\sqrt{
\frac{
\displaystyle \iint_{xz} |E_{\mathrm{opt}}(x,z)|^2 \, dx\,dz
}{
S_{\mathrm{opt}}
}
}.
\end{equation}
for optical modes (including 2 pumps, 1 probe and optical sidebands). 

The corresponding normalized mode profiles are then defined as
\begin{equation}
g_{\Omega}(x,z)
=
\frac{E_{\Omega}(x,z)}{A_{\Omega}},
\end{equation}
and
\begin{equation}
g_{\mathrm{opt}}(x,z)
=
\frac{E_{\mathrm{opt}}(x,z)}{A_{\mathrm{opt}}}.
\end{equation}

With this normalization, the total normalized mode powers satisfy
\begin{equation}
\iint_{xz} |g_{\Omega}(x,z)|^2 \, dx\,dz = S_{\Omega},
\end{equation}
and
\begin{equation}
\iint_{xz} |g_{\mathrm{opt}}(x,z)|^2 \, dx\,dz = S_{\mathrm{opt}}.
\end{equation}

The electro-optic overlap factor describes the microwave field sampled by the optical mode intensity inside the lithium-niobate interaction region. It is defined as
\begin{equation}\label{Eq_gamma_eo}
\Gamma_{\mathrm{eo}}
=
\frac{
\displaystyle \iint_{\mathrm{LN}}
\left| g_{\mathrm{opt}}(x,z) \right|^2
g_{\Omega}(x,z) \, dx\,dz
}{
\displaystyle \iint_{xz}
\left| g_{\mathrm{opt}}(x,z) \right|^2 \, dx\,dz
}.
\end{equation}

Equivalently, since the denominator is equal to \(S_{\mathrm{opt}}\), this can be written as
\begin{equation}
\Gamma_{\mathrm{eo}}
=
\frac{1}{S_{\mathrm{opt}}}
\displaystyle \iint_{\mathrm{LN}}
\left| g_{\mathrm{opt}}(x,z) \right|^2
g_{\Omega}(x,z) \, dx\,dz .
\end{equation}

The reciprocal optical-to-electrical overlap factor describes the projection of the optically induced source term onto the microwave mode. It uses the same overlap numerator but is normalized by the microwave mode power:
\begin{equation}\label{Eq_gamma_oe}
\Gamma_{\mathrm{oe}}
=
\frac{
\displaystyle \iint_{\mathrm{LN}}
\left| g_{\mathrm{opt}}(x,z) \right|^2
g_{\Omega}(x,z) \, dx\,dz
}{
\displaystyle \iint_{xz}
\left| g_{\Omega}(x,z) \right|^2 \, dx\,dz
}.
\end{equation}

Since the denominator is equal to \(S_{\Omega}\), the optical-to-electrical overlap can also be written as
\begin{equation}
\Gamma_{\mathrm{oe}}
=
\frac{1}{S_{\Omega}}
\displaystyle \iint_{\mathrm{LN}}
\left| g_{\mathrm{opt}}(x,z) \right|^2
g_{\Omega}(x,z) \, dx\,dz .
\end{equation}

The mmWave-THz mode profiles are simulated using CST Studio, as illustrated in Supplementary Fig.~\ref{FigS_A6}a, with the device dimensions summarized in Supplementary Table~\ref{Table_dim}. The simulated $x$-component electric-field profiles are shown in Supplementary Fig.~\ref{FigS_A6}b-d for the optical mode at 1550~nm, the mmWave mode at 180~GHz, and the terahertz mode at 2~THz, respectively. For the same waveguide and electrode geometry, at both 180~GHz and 2~THz, the mmWave-THz fields remain well confined within the electrode gap, with only slight changes in their spatial distributions.
Using Eqs.~(\ref{Eq_S_Om}) and~(\ref{Eq_S_opt}), we estimate an optical mode area of $S_\mathrm{opt}\approx1~\mu\mathrm{m}^2$ and mmWave-THz  mode areas of $S_\Omega\approx31~\mu\mathrm{m}^2$ at 180~GHz and $S_\Omega\approx29~\mu\mathrm{m}^2$ at 2~THz. Accordingly, the electro-optic mode-overlap factor $\Gamma_{\mathrm{eo}}$ and the optical-to-electrical mode-overlap factor $\Gamma_{\mathrm{oe}}$ vary only weakly across themmWave-THz  range. Using Eqs.~(\ref{Eq_gamma_eo}) and~(\ref{Eq_gamma_oe}), we estimate $\Gamma_{\mathrm{eo}}\approx0.6$ and $\Gamma_{\mathrm{oe}}\approx0.02$.

\newpage

\subsection{Photon-flux conversion efficiencies for mmWave-THz generation and detection}


\begin{figure}[h!]
\centering
    \includegraphics[width=\textwidth]{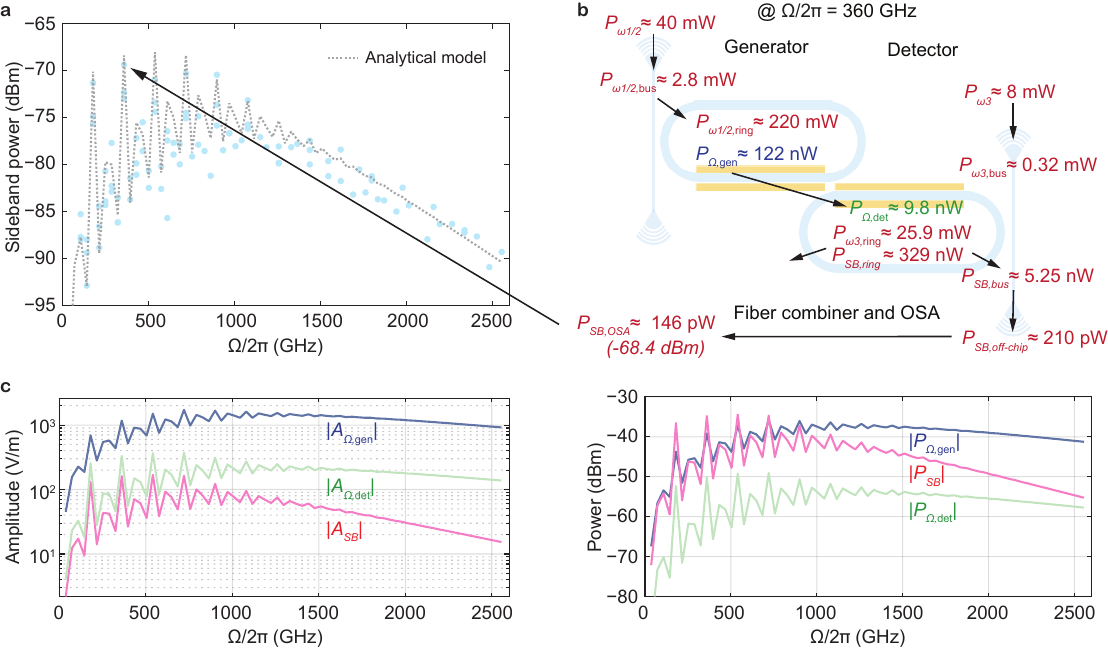}
    \caption{\textbf{Photon-flux calibration and extraction of conversion efficiencies.}
    \textbf{a}. Model fit to the measured frequency-dependent response of the signal core. Blue points show the off-chip optical sideband power measured using an optical spectrum analyzer (OSA), and solid curves show the corresponding theoretical results.
    \textbf{b}. Calibrated power flow through the generation and detection processes, including the input optical pump and probe powers, the generated and transferred on-chip mmWave–THz powers, the intracavity optical probe and sideband powers, and the final off-chip sideband power measured by the OSA. Here the power flow at 360 GHz is shown as a representative example.
    \textbf{c}. 
    On-chip mmWave–THz electric-field amplitudes and powers. $A_\mathrm{\Omega,gen}$ and $P_\mathrm{\Omega,gen}$ denote the mmWave–THz electric-field amplitude and power, respectively, in the generator CPS cavity;
    $A_\mathrm{\Omega,det}$ and $P_\mathrm{\Omega,det}$ denote the transferred mmWave–THz electric-field amplitude and power, respectively, in the detector CPS cavity; 
    and $A_\mathrm{SB}$ and $P_\mathrm{SB}$ denote the upconverted optical sideband field amplitude and power, respectively, in the detector racetrack resonator.
    }
\label{FigS_A7}
\end{figure}

We next combine the theoretical models, numerical simulations, and experimental characterizations described above to quantify the photon-flux conversion efficiencies of the signal core. 
The generated on-chip mmWave-THz field, including CPS-cavity enhancement, is calculated from Eq.~(\ref{Eq_gen}), while the field transferred to the detection module is calculated from Eq.~(\ref{Eq_det}). 
These fields are converted to the corresponding powers, $P_{\Omega,\mathrm{gen}}$ and $P_{\Omega,\mathrm{det}}$, using Eq.~(\ref{Eq_THzPower}). In the detection module, the optical sideband field $A_\mathrm{SB}$ and sideband power $P_\mathrm{SB}$ are obtained from Eqs.~(\ref{Eq_A_SB}) and~(\ref{Eq_P_SB}), respectively.

We then combine the model with previously extracted parameters, such as the grating-coupler efficiency $\eta$ and electro-optic mode-overlap factors $\Gamma$, to fit the measured frequency-dependent response of the signal core.
Representative parameters are listed in Supplementary Table~\ref{Table1}, and the complete fitting code and parameter set are available in the accompanying Zenodo archive. 
As shown in Supplementary Fig.~\ref{FigS_A7}a, the model reproduces the measured off-chip sideband power represented by the blue data points. 
The analytical model captures the overall frequency dependence well.
The residual variations among datasets are attributed mainly to day-to-day changes in the resonance alignment and stability of the racetrack cavities, which are sensitive to variations in ambient temperature and humidity.

\begin{table}[h]
\centering
\caption{Representative parameters in generation-detection model.}
\renewcommand{\arraystretch}{0.8}
\setlength{\tabcolsep}{5pt}

\begin{tabularx}{0.75\linewidth}{@{}p{0.1\linewidth} p{0.22\linewidth} X@{}}
\toprule
\textbf{Symbol} & \textbf{Value} & \textbf{Explanation / notes} \\
\midrule

$\chi^{(2)}$
& \SI{360}{\pico\meter\per\volt}
& Second-order nonlinear coefficient. \\

$\Omega$
& $2\pi \times \SI{360} {\mathrm{GHz}}$
& mmWave-THz angular frequency. \\

$n_\Omega$
& $2.23$
& THz refractive index. \\

$c_0$
& \SI{3e8}{\meter\per\second}
& Speed of light in vacuum. \\

$\varepsilon_0$
& \SI{8.854e-12}{\farad\per\meter}
& Vacuum permittivity. \\

$n_{\mathrm{eff}}$
& $1.91$
& Effective optical refractive index. \\

$P_\mathrm{o1,o2}$
& \SI{250}{\milli\watt}
& Intracavity optical pump power. \\

$P_{\mathrm{o3}}$
& \SI{25}{\milli\watt}
& Intracavity optical probe power. \\

$\alpha$
& \SI{3.3}{dB\per\milli\meter}
& CPS power-loss coefficient at 360~GHz. \\

$L_\mathrm{cps}$
& \SI{750e-6}{\meter}
& CPS / interaction length. \\

$\varepsilon_r$
& $1.91^2$
& Relative permittivity for lossless optics. \\

$\omega_\mathrm{SB}$
& $2\pi \times \SI{194e12}{\hertz}$
& Optical sideband angular frequency. \\

$\eta_\mathrm{gen}$
& $7\%$
& Grating coupler efficiency in generator. \\

$\eta_\mathrm{det}$
& $4\%$
& Grating coupler efficiency in detector. \\

$\Gamma_\mathrm{eo}$
& $0.6$
& Electrical-to-optical modal overlap factor. \\

$\Gamma_\mathrm{oe}$
& $0.02$
& Optical-to-electrical modal overlap factor. \\

$\kappa$
& $0.15$
& field coupling coefficient between CPS. \\

$r_{1/2}$
& $0.9$
& Reflection coefficient. \\

$S_{\Omega}$
& \SI{30e-12}{\meter\squared}
& Effective mmWave-THz mode area. \\

$S_{\mathrm{opt}}$
& \SI{1e-12}{\meter\squared}
& Effective optical mode area. \\

\bottomrule
\label{Table1}
\end{tabularx}
\end{table}

Supplementary Fig.~\ref{FigS_A7}b and supplementary Table~\ref{Table_PowerFlow_RoomT} summarize the on-chip power flow when measured at 360~GHz as an example. 
On the generation side, the two pump lasers each have an output power of approximately 12 mW.

Their outputs are combined using a 50:50 fiber coupler, resulting in a total power of approximately 10~mW at the input of the erbium-doped fiber amplifier. 
The amplifier produces a total output power of approximately 100~mW across the two optical tones.
After accounting for an optical path loss of approximately 20\%, the off-chip pump powers at the device input are $P_{\omega_1}\approx P_{\omega_2}\approx40$~mW.
With a grating-coupler efficiency of approximately 7\%, the on-chip bus-waveguide powers are $P_{\omega_\mathrm{1,bus}}\approx P_\mathrm{\omega_2,bus}\approx 2.8$~mW.
A power-buildup factor of approximately 81 results in circulating powers of $P_{\omega_\mathrm{1,ring}}\approx P_\mathrm{\omega_2,ring}\approx 220$~mW inside the racetrack resonator (see Supplementary Information~\ref{SI_theory}.5). 
The mmWave–THz power generated in the CPS cavity is approximately 122~nW calculated by using Eqs.~(\ref{Eq_gen}) and (\ref{Eq_THzPower}), and the power transferred to the detection module is $P_{\Omega,\mathrm{det}} \approx9.8$~nW calculated via Eqs.~(\ref{Eq_det}) and (\ref{Eq_THzPower}). 

On the detection side, the racetrack resonator is probed by a third laser with an output power of 10~mW.
After accounting for an optical path loss of approximately 20\%, the off-chip probe power at the device input is $P_{\omega_3}\approx8$~mW. 
The grating-coupler efficiency on the detector side is approximately 4\%, resulting in an on-chip probe power of $P_{\omega_\mathrm{3,bus}}\approx 0.32$~mW. 
With a power-buildup factor of approximately 81, the calculated circulating probe power inside the racetrack resonator is $P_{\omega_\mathrm{3,ring}}\approx25.9$~mW. 
Using Eqs.~(\ref{Eq_A_SB}) and (\ref{Eq_P_SB}), the circulating power of the upconverted sideband inside the detector racetrack resonator is $P_\mathrm{SB,ring}\approx329$~nW when the intracavity power-enhancement factor of approximately 5000 for an internally generated field is included (see Supplementary Information~\ref{SI_theory}.5).
The sideband power coupled into the bus waveguide is $P_\mathrm{SB,bus}=P_\mathrm{SB,ring}\cdot(1-r^2)$, where the field reflection coefficient $r\approx0.992$. 
We therefore obtain $P_\mathrm{SB,bus}\approx5.25$~nW. 
After accounting for the grating-coupler efficiency of 4\%, the off-chip sideband power is $P_\mathrm{SB,off-chip}\approx210$~pW.
An additional insertion loss of approximately 30\% before the optical spectrum analyzer results in a measured power of $P_{\mathrm{SB,OSA}}\approx146$~pW, consistent with the measurements as shown in Supplementary Fig.~\ref{FigS_A7}a.

Together, the model fitting and calibrated power-flow analysis shown in Supplementary Fig.~\ref{FigS_A7}a,b validate the power-calibration procedure and allow us to infer the on-chip mmWave-THz electric-field amplitudes and powers, as shown in Supplementary Fig.~\ref{FigS_A7}c.

\begin{table}[h]
\centering
\caption{\textbf{Power flow on chip operated at 360~GHz.}}
\label{Table_PowerFlow_RoomT}
\begin{tabular}{lll}
\hline
Parameter & Symbol & Value \\
\hline
Laser pump power (off-chip) & $P_{\omega1}$ and $P_{\omega2}$ & $40~\mathrm{mW}$ \\
Laser pump power (in bus waveguide) & $P_\mathrm{\omega1,bus}$ and $P_\mathrm{\omega2,bus}$ & $2.8~\mathrm{mW}$ \\
Intracavity pump power of generator ring & $P_\mathrm{\omega1,ring}$ and $P_\mathrm{\omega2,ring}$ & $220~\mathrm{mW}$ \\
Generated mmWave-THz power in generator CPS & $P_\mathrm{\Omega,gen}$ & $122~\mathrm{nW}$ \\
Transferred mmWave-THz power in detector CPS & $P_\mathrm{\Omega,det}$ & $9.8~\mathrm{nW}$ \\
Laser probe power (off-chip) & $P_{\omega3}$ & $8~\mathrm{mW}$ \\
Intracavity probe power of detector ring & $P_\mathrm{\omega3,ring}$ & $25.9~\mathrm{mW}$ \\
Intracavity sideband power of detector ring & $P_\mathrm{SB,ring}$ & $329~\mathrm{nW}$ \\
On-chip sideband power in bus waveguide & $P_\mathrm{SB,bus}$ & $5.25~\mathrm{nW}$ \\
Off-chip sideband power & $P_\mathrm{SB,off-chip}$ & $210~\mathrm{pW}$ \\
Measured sideband power in OSA & $P_\mathrm{SB,OSA}$ & $146~\mathrm{pW}$ \\

\hline
\end{tabular}
\end{table}

Using these calibrated on-chip powers, we define the pump-power-normalized photon-flux generation efficiency as the ratio of the generated mmWave–THz photon flux to the photon flux of the $\omega_1$ optical tone, normalized by the on-chip power of the $\omega_2$ tone:
\begin{equation}\label{Eq_Eta_gen}
\eta_\mathrm{gen}=
\frac{P_{\Omega,\mathrm{gen}}/\hbar\Omega}
{P_{\omega_2,\mathrm{bus}}/\hbar\omega_2
\cdot P_{\omega_1,\mathrm{bus}}}
=\frac{P_{\Omega,\mathrm{gen}}}
{P_{\omega_2,\mathrm{bus}}
\cdot P_{\omega_1,\mathrm{bus}}}
\cdot \frac{\omega_2}{\Omega}.
\end{equation}

In our measurements, the two on-chip pump powers are nearly equal such that $P_{\omega_1,\mathrm{bus}}\approx P_{\omega_2,\mathrm{bus}}$). The two optical frequencies are also very close, satisfying $(\omega_1-\omega_2)/[\frac{1}{2}(\omega_1+\omega_2)]<1\%$. We therefore neglect the small difference between the two optical pump frequencies when evaluating the normalized efficiency.

Similarly, we define the pump-power-normalized photon-flux detection efficiency as the ratio of the upconverted optical sideband photon flux to the incident mmWave–THz photon flux, normalized by the on-chip optical probe power:
\begin{equation}\label{Eq_Eta_det}
\eta_\mathrm{det}=
\frac{P_\mathrm{SB}/\hbar\omega_\mathrm{SB}}
{P_{\Omega,\mathrm{det}}/\hbar\Omega
\cdot P_{\omega_3,\mathrm{bus}}}
=\frac{P_\mathrm{SB,bus}}
{P_{\Omega,\mathrm{det}}
\cdot P_{\omega_3,\mathrm{bus}}}
\cdot \frac{\Omega}{\omega_\mathrm{SB}}
\end{equation}

These results are reported in Fig.3e of the main text, where we find maximal efficiencies reaching $\eta_\mathrm{gen}(720\mathrm{GHz})=1.02\%/\mathrm{mW}$ and $\eta_\mathrm{det}(864\mathrm{GHz})=0.52\%/\mathrm{mW}$. 
For the on-chip pump powers used in our experiments, $P_\mathrm{\omega2,bus}=2.8$~mW and $P_\mathrm{\omega3,bus}=0.32$~mW, these normalized peak values correspond to total photon-flux conversion efficiencies of 2.86\% for generation and 0.17\% for detection.


\newpage
\section{Signal-core-enabled photonic probing of mmWave-THz loss on chip}\label{SI_loss}

\begin{figure}[h]
    \centering
    \includegraphics[width=1\textwidth]{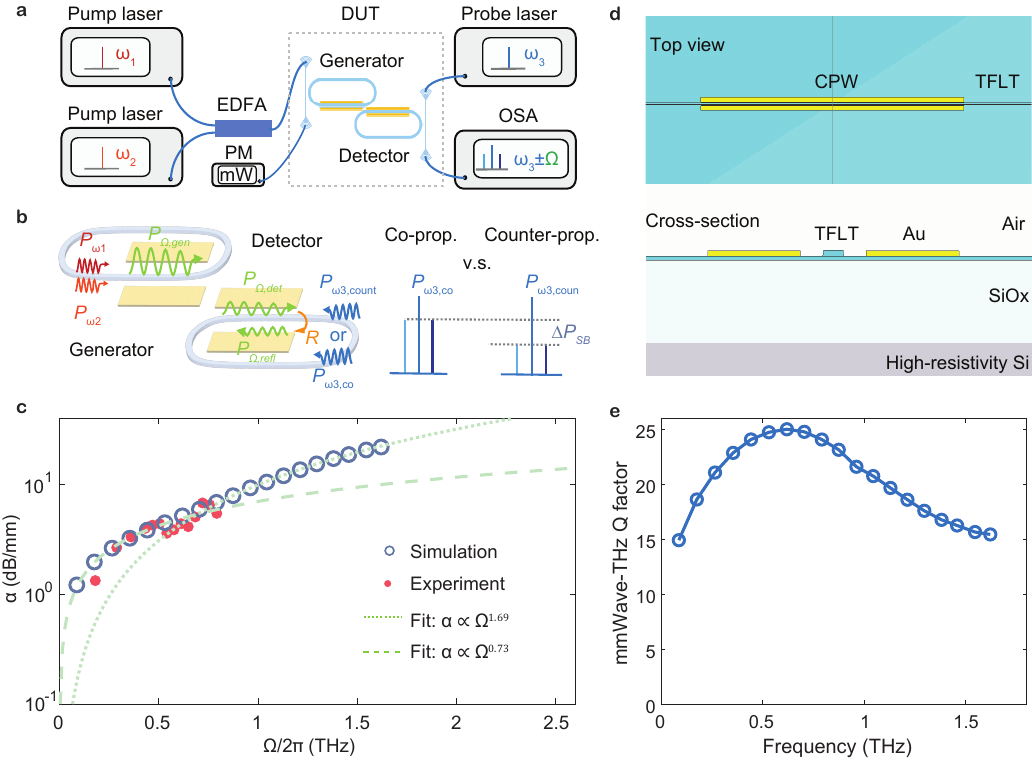}
    \caption{
    \textbf{Signal-core-enabled photonic probing of on-chip mmWave-THz loss.}
    \textbf{a}. Experimental setup for on-chip characterization of electro-optic
    conversion efficiency and mmWave-THz loss using the signal-core architecture.
    \textbf{b}. Principle of photonic probing of on-chip mmWave--THz loss.
    Co-propagating optical and mmWave--THz fields produce efficient phase
    modulation, whereas counter-propagating fields are phase-mismatched. The
    resulting contrast $\Delta P_{\mathrm{SB}}$ in the optical sideband power
    encodes the on-chip loss as $\Delta P_{\mathrm{SB}} \propto
    e^{-\alpha L_{\mathrm{cps}}}$.
    \textbf{c}. Extracted loss $\alpha$ from experiments (orange dots), simulations
    (blue circles), and analytical fits (green curves). Two regimes are observed:
    conductive-loss-dominated behavior at lower frequencies (dotted fit) and
    radiation-loss-dominated behavior at higher frequencies (dashed fit).
    \textbf{d}. CST model of the CPS cavity used to corroborate $\alpha$, with
    dimensions identical to those of the fabricated devices. 
    \textbf{e}. Q factors of mmWave-THz modes calculated using CST Studio. 
    }
    \label{FigS_B}
\end{figure}

On-chip mmWave-THz loss is a key limiting factor for device performance, but conventional electronic characterization is constrained by instrument bandwidth and by the difficulty of accessing micrometer-scale structures on chip. We therefore develop an in situ photonic probing method~\cite{gaier2025wireless}, based on the signal-core architecture, to measure the on-chip mmWave-THz propagation loss.

Supplementary Fig.~\ref{FigS_B}a shows the experimental setup, and Supplementary Fig.~\ref{FigS_B}b illustrates the probing concept based on traveling-wave phase matching. The mmWave-THz field generated in the generator module has power $P_{\Omega,\mathrm{gen}}$ and co-propagates with the optical pumps from left to right. It is then coupled into the detector module with power $P_{\Omega,\mathrm{det}}$ and reaches the CPS termination on the right side of the device. After reflection, it propagates back through the detector module with power $P_{\Omega,\mathrm{refl}}\approx P_{\Omega,\mathrm{det}} \cdot R \cdot e^{-\alpha L_\mathrm{cps}}$, where $R$ is the power reflection coefficient and $L_\mathrm{cps}$ is the CPS propagation length.

When the optical probe $P_{\omega_3,\mathrm{co}}$ co-propagates with the mmWave-THz field in the detector CPS cavity, traveling-wave phase matching produces strong optical sidebands. In the counter-propagating configuration, $P_{\omega_3,\mathrm{count}}$, the forward-propagating mmWave-THz field is phase mismatched with the optical probe and therefore produces negligible modulation. Efficient modulation occurs only after the mmWave-THz field reflects from the termination and co-propagates with the counter-propagating optical probe. The counter-propagating measurement therefore includes an additional single-pass propagation energy loss term $e^{-\alpha L_\mathrm{cps}}$. Consequently, the sideband-power reduction $\Delta P_\mathrm{SB}$ between the co- and counter-propagating configurations encodes the mmWave-THz propagation loss and allows extraction of the effective loss coefficient $\alpha$.

In this analysis, we assume $R\approx1$ and absorb parasitic effects, including fringing-field radiation, mode conversion, and impedance mismatch, into the effective loss coefficient $\alpha$. We perform the measurement from 100~GHz to approximately 800~GHz, with the upper frequency limited by the signal-to-noise ratio. The red dots in Supplementary Fig.~\ref{FigS_B}c shows the extracted loss measured on the signal core.

We also use CST Studio Suite to calculate the eigenmodes of the CPS cavity structure, whose model is shown in Supplementary Fig.~\ref{FigS_B}d with cross-section dimensions same as Supplementary Table~\ref{Table_dim}. Here, to compare the measured mmWave-THz loss with electromagnetic simulations, we extracted $\alpha$ from the simulated quality factors of the CPS cavity modes. The stored electromagnetic energy decays along the propagation direction as
\begin{equation}
    U(z) = U_0 e^{-\alpha z},
\end{equation}
where $\alpha$ is in unit $\mathrm{m^{-1}}$. The internal quality factor is related to the energy decay rate by
\begin{equation}
    Q_m = \frac{\omega_m}{\alpha_m v},
\end{equation}
where $\omega_m = 2\pi f_m$, with $f_m$ denoting the resonance frequency of the $m$-th cavity mode, and $v$ denoting the phase velocity of the CPS mode. Thus,
\begin{equation}
    \alpha_m
    =
    \frac{2\pi f_m}{Q_m v}.
\end{equation}

Or we can express $\alpha$ in the unit of $\mathrm{dB/m}$ as
\begin{equation}
    \alpha_{\mathrm{m,dB}}
    =
    4.343 \times
    \frac{2\pi f_m}{Q_m v}.
    \label{eq:alpha_from_Q_energy}
\end{equation}

For the half-wavelength CPS cavity, the phase velocity is estimated from the
fundamental resonance frequency $f_1$ as $ v = 2L f_1$, where $L$ is the cavity length. Substituting this relation into
Eq.~\eqref{eq:alpha_from_Q_energy}, the propagation loss at each measured or
simulated resonance frequency is obtained as
\begin{equation}\label{Eq_mmW_alpha}
    \alpha_{\mathrm{m,dB}}
    =
    4.343 \times
    \frac{\pi m}{L Q_m}.
\end{equation}

The corresponding mmWave-THz mode $Q$ factors from CST simulations are shown in Supplementary Fig.~\ref{FigS_B}e. From these values, we extract the propagation loss using Eq.~(\ref{Eq_mmW_alpha}), shown by the blue circles in Supplementary Fig.~\ref{FigS_B}c. The simulated loss agrees well with the experimentally extracted values, shown by the red dots. The total loss is dominated by conductor loss in the gold CPSs, $\alpha_{\mathrm{cond}}$, and radiation loss, $\alpha_{\mathrm{rad}}$. The extracted frequency dependence follows $\alpha \propto f^{0.73}$ at lower frequencies and $\alpha \propto f^{1.69}$ at higher frequencies, indicating a crossover from predominantly conductor-loss-limited behaviour, for which $\alpha_{\mathrm{cond}} \propto f^{0.5}$~\cite{keil2002high}, toward a regime with an increasing radiation-loss contribution, where $\alpha_{\mathrm{rad}} \propto f^{3}$~\cite{frankel2002terahertz}.

\newpage

\section{RF-compatible photonic platform for CPS-cavity study and extraction of $\kappa$ and $\alpha$} \label{SI_LossRF}

\begin{figure}[h!]
    \centering
    \includegraphics[width=1\textwidth]{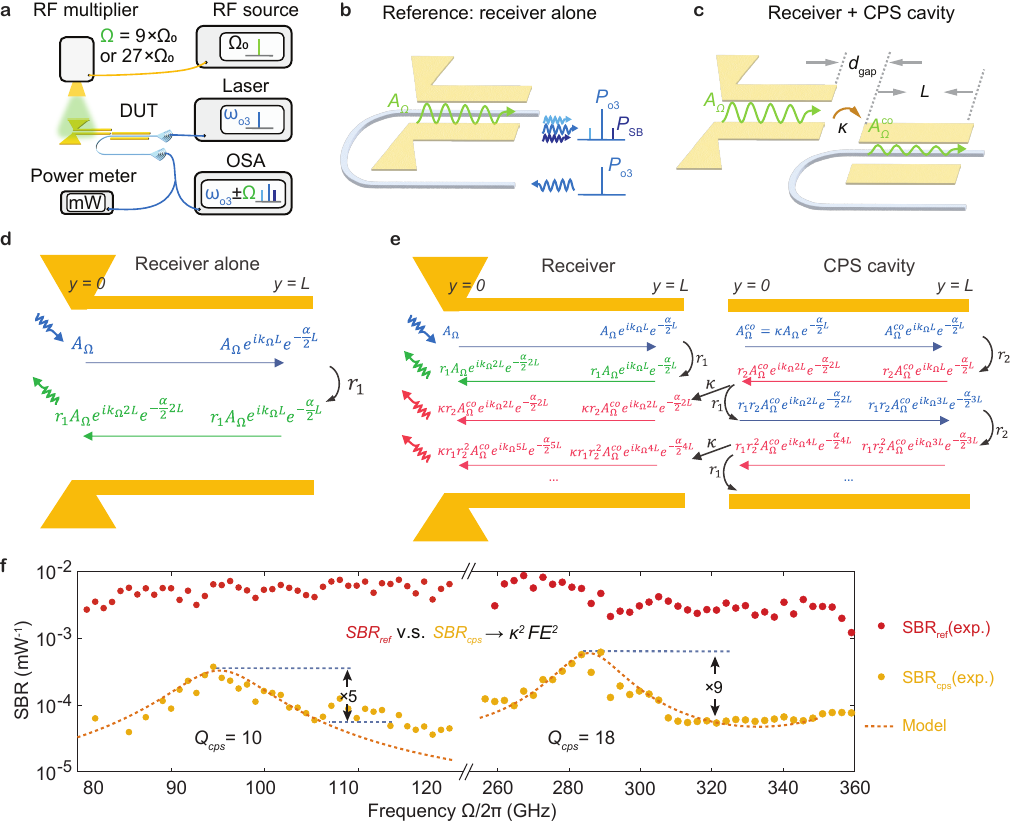}
    \renewcommand{\baselinestretch}{1} 
    \caption{\textbf{RF-compatible photonic characterization of CPS cavity responses.}
    \textbf{a}. Experimental setup for RF-assisted electro-optic characterization of mmWave-THz CPS structures. A commercial RF multiplier radiates calibrated mmWave signals onto the chip, where they are captured by a bowtie antenna and launched into the CPS structure. The mmWave field modulates an optical probe, generating sidebands that are detected optically.
    \textbf{b}. Schematic of the broadband antenna-CPS reference receiver. The bowtie antenna captures the radiated mmWave signal and launches it into the CPS, where the field is read out by electro-optic sideband generation from an optical probe.
    \textbf{c}. Schematic of the cavity-coupled device under test, in which the antenna-CPS receiver capacitively couples the captured mmWave field into a separate CPS cavity with coupling coefficient $\kappa$.
    \textbf{d}. Field-evolution model for the reference receiver, including propagation phase and loss-induced amplitude attenuation. Multiple round trips are suppressed because the returning field is predominantly re-radiated into free space by the bowtie antenna.
    \textbf{e}. Fabry-Perot-type model for the cavity-coupled device, where coherent summation of multiple round trips in the CPS cavity produces resonant field enhancement.
    \textbf{f}. Measured normalized sideband response of the reference receiver and cavity-coupled device. The reference receiver shows a nearly flat broadband response, whereas the cavity-coupled device exhibits resonances near 95~GHz and 285~GHz with response enhancements of approximately 5 and 9, respectively. Dashed curves show fits based on the CPS-cavity response model, yielding $\alpha\sim1.7$~dB/mm and $\kappa\sim0.07$ near 100~GHz, and $\alpha\sim3$~dB/mm and $\kappa\sim0.12$ near 300~GHz.
    }
    \label{FigS_C}
\end{figure}

As a complementary experiment to extract the on-chip coupling coefficient $\kappa$ and loss coefficient $\alpha$, we develop an RF-compatible photonic platform for characterizing mmWave-THz CPS cavity responses, as shown in Supplementary Fig.~\ref{FigS_C}a. The platform uses a commercial RF multiplier as a calibrated free-space mmWave source, providing output powers of approximately 100~mW in the 80-120~GHz band and approximately 4~mW in the 260-380~GHz band. To couple the free-space mmWave signal into the on-chip CPS structure for electro-optic readout, we integrate a bowtie antenna co-designed with the CPS to form a broadband receiver, as shown in Supplementary Fig.~\ref{FigS_C}b. This antenna-CPS receiver acts as an on-chip injection port that launches the captured mmWave field into a separate CPS section, which serves as the cavity under test as shown in Supplementary Fig.~\ref{FigS_C}c. Unlike the signal-core experiment, where two CPS cavities contribute to the measured response, this antenna-coupled platform isolates a single CPS cavity and therefore enables unambiguous extraction of its intrinsic response.

The radiated mmWave signal is collected by the bowtie antenna and launched into the CPS with field amplitude $A_\Omega$ and corresponding power $P_\Omega$. The CPS response is read out through electro-optic sideband conversion: the mmWave field modulates an optical probe with power $P_\mathrm{o3}$, generating optical sidebands at $\omega_3\pm\Omega$ with power $P_\mathrm{SB}\propto P_\Omega P_\mathrm{o3}$. We therefore define the normalized sideband response as
\begin{equation}
{SBR}=\frac{P_\mathrm{SB}}{P_\Omega P_{\omega_3}},
\end{equation}
which serves as the figure of merit for the power-normalized CPS response.

We first analyze the response of the non-resonant reference receiver, which does not include an additional CPS cavity. Supplementary Fig.~\ref{FigS_C}d illustrates the evolution of the mmWave field along the receiver, including the accumulated propagation phase and amplitude attenuation associated with the loss coefficient $\alpha$. After reaching the right termination, the mmWave field is partially reflected and propagates back toward the bowtie antenna. Because the antenna-CPS receiver is designed to couple efficiently to free space, most of the returning field is re-radiated rather than reflected back into the CPS. Together with the relatively large propagation loss of the gold CPS, this suppresses multiple round trips and makes the cavity effect negligible in the reference structure.

We experimentally confirm this behaviour by illuminating the reference receiver with the calibrated mmWave source across the measured frequency bands. The up-converted sideband signal exhibits a nearly flat normalized response, as shown by the red data points in Supplementary Fig.~\ref{FigS_C}f, confirming that the antenna-CPS receiver provides a broadband reference response without pronounced cavity enhancement. In the absence of CPS-cavity resonant effects, the reference response can be expressed using Eq.~(\ref{Eq_SFG_simple}) as
\begin{equation}\label{Eq_ref}
{SBR}_\mathrm{ref}
=\frac{|A_\mathrm{SB}|^2}{|A_\mathrm{o3}|^2 |A_{\Omega}^\mathrm{co}|^2}
=\frac{\left(\chi^{(2)}\omega_\mathrm{SB}\right)^2}{n^2 c_0^2 \alpha^2}
\Gamma_\mathrm{eo}^2
\left(1-e^{-\alpha L/2}\right)^2 .
\end{equation}
Here, $A_\mathrm{SB}$ is the amplitude of the up-converted sideband field, $A_\mathrm{o3}$ is the amplitude of the optical probe, and $A_{\Omega}^\mathrm{co}=A_{\Omega}$ is the amplitude of the mmWave-THz field co-propagating with the optical probe.

In contrast, the device under test incorporates a CPS cavity capacitively coupled to the antenna-CPS receiver with coupling coefficient $\kappa$, as shown in Supplementary Fig.~\ref{FigS_C}c. The measured response of this cavity-coupled device is shown by the yellow data points in Supplementary Fig.~\ref{FigS_C}f. Compared with the reference response, shown by the red data points, the cavity-coupled device exhibits two key features. First, the overall response baseline is lower because only a fraction of the mmWave field collected by the receiver is capacitively transferred into the CPS cavity under test. Second, pronounced resonant peaks appear near 95~GHz and 285~GHz, enhancing the normalized response by factors of approximately 5 and 9, respectively. The 95~GHz resonance corresponds to the fundamental CPS-cavity mode and is slightly higher in frequency than that of the coupled CPS cavities in the signal core, primarily because the device studied here uses a silica cladding layer, whereas the signal core uses an air-clad CPS structure. Compared with the nearly flat response of the non-resonant reference receiver, these peaks confirm that the measured enhancement originates from the CPS-cavity resonance rather than from the antenna or free-space coupling response.

To extract $\kappa$ and $\alpha$, we develop a Fabry-Perot-type model for the cavity-coupled device, as illustrated in Supplementary Fig.~\ref{FigS_C}e. The model tracks the mmWave fields at the left and right terminations of the receiver and CPS cavity, including propagation phase and amplitude attenuation induced by the loss coefficient $\alpha$. As in the reference receiver, we assume that right-to-left propagating fields in the receiver section are predominantly emitted back into free space by the impedance-matched bowtie antenna, so that back-reflection into the CPS is negligible. In the CPS cavity section, the co-propagating field from left to right is enhanced by the coherent summation of multiple round trips and can be written as
\begin{equation}\label{Eq_CPS_cavity}
A_{\Omega}^{\mathrm{co,total}}
= A_{\Omega}^{\mathrm{co}}\left(1+ r_1 r_2 e^{i k_{\Omega}2L} e^{-\frac{\alpha}{2}2L}+ r_1^2 r_2^2 e^{i k_{\Omega}4L} e^{-\frac{\alpha}{2}4L}+\cdots\right)
=\frac{A_{\Omega}^{\mathrm{co}}}{1 - r_1 r_2 e^{i k_{\Omega}2L} e^{-\alpha L}}
\end{equation}
where $A_{\Omega}^{\mathrm{co}}=\kappa A_{\Omega} e^{-\frac{\alpha}{2}L}$ is the mmWave field amplitude transferred from the right termination of the receiver to the left termination of the CPS cavity. The geometric series represents the CPS-cavity field-enhancement factor. Considering the field enhancement in Eq.~(\ref{Eq_CPS_cavity}) together with Eq.~(\ref{Eq_SFG_simple}), the normalized sideband response of the CPS cavity can be written as
\begin{equation}
\begin{aligned}
{SBR_\mathrm{cps}}
&=
\left(
\frac{\chi^{(2)}\omega_\mathrm{SB}}{n c_0\alpha}
\right)^2
\Gamma_\mathrm{eo}^2
\left(
1-e^{-\alpha L/2}
\right)^2
|\kappa|^2
e^{-\alpha L}
\left|
\frac{1}{
1-r_1r_2
e^{-\alpha L}
e^{i2k_{\Omega}L}
}
\right|^2 .
\end{aligned}
\end{equation}

Therefore, direct comparison between the reference response and the CPS-cavity-coupled response enables extraction of $\kappa$ and $\alpha$:
\begin{equation}\label{Eq_FPFit} 
SBR_\mathrm{cps} = \frac{ |\kappa|^2 e^{-\alpha L} }{ \left| 1-r_1r_2 e^{-\alpha L} e^{i2k_{\Omega}L} \right|^2 } \cdot SBR_\mathrm{ref} 
\end{equation}

We then use Eq.~(\ref{Eq_FPFit}) to fit the experimental response of the CPS-cavity device, shown by the yellow data points and dashed curves in Supplementary Fig.~\ref{FigS_C}f. The fits yield loss coefficients of $\alpha\sim1.8$~dB/mm and $\alpha\sim3$~dB/mm, with coupling coefficients of $\kappa\sim0.07$ and $\kappa\sim0.12$, near 100~GHz and 300~GHz, respectively.

\begin{figure}[h]
    \centering
    \includegraphics[width=\textwidth]{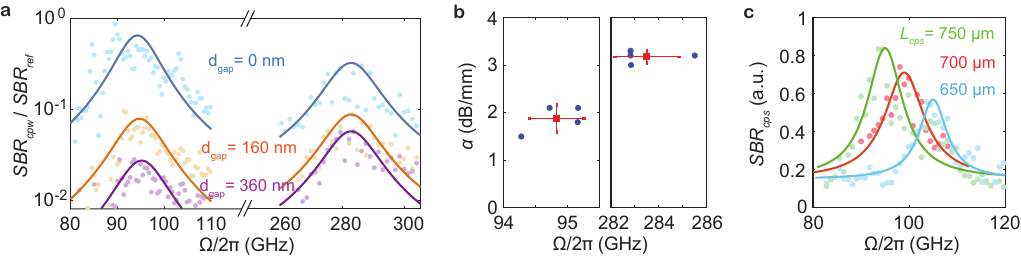}
    \renewcommand{\baselinestretch}{1} 
    \caption{\textbf{RF-compatible photonic platform for CPS-cavity characterization.}
    \textbf{a}. Power-coupling efficiency $\mathrm{SBR_{cps}/SBR_{ref}}$ as a function of CPS gap $d_\mathrm{gap}$, measured near 100~GHz and 300~GHz. When the two CPSs are in contact ($d_\mathrm{gap}=0$), the coupling efficiency is approximately 0.7 near 100~GHz and 0.6 near 300~GHz. The efficiency decreases to approximately 0.1 when $d_\mathrm{gap}=160$~nm.
    \textbf{c}.Fundamental CPS resonance frequency as a function of cavity length $L_\mathrm{cps}$, showing the tunability of the CPS resonance and the ability to engineer commensurability between the CPS and optical resonances.
    }
    \label{FigS_C2}
\end{figure}

We further fabricated another batch of sample with various $d_\mathrm{gap}$ to study the capacitive coupling coefficient $\kappa$ as it plays a central role in efficient on-chip signal routing and transmission between individual signal cores, and is primarily controlled by the gap size $d_\mathrm{gap}$. Supplementary Fig.~\ref{FigS_C2}a shows the measured power-coupling efficiency from the receiver to the CPS cavity, defined as $SBR_\mathrm{cps}/SBR_\mathrm{ref}$. When the two CPSs are in contact, corresponding to $d_\mathrm{gap}=0$, the measured efficiency is approximately 0.7 near 100~GHz and 0.6 near 300~GHz. As the gap increases, the coupling efficiency decreases, reaching approximately 0.1 at $d_\mathrm{gap}=160$~nm, which is very close to the configuration of the signal core ($d_\mathrm{gap}\approx200$~nm). The corresponding coupling coefficients $\kappa$ are summarized in Fig.~2f of the main text.

We further use Eq.~(\ref{Eq_FPFit}) to fit the measured responses in Supplementary Fig.~\ref{FigS_C2}a and extract the effective loss coefficient $\alpha$. For the fabricated devices shown in Supplementary Fig.~\ref{FigS_C}f and Fig.~\ref{FigS_C2}a, the fitted fundamental resonance has a mean frequency of $94.8\pm0.4$~GHz and a loss of $1.87\pm0.29$~dB/mm, and the higher-order mode has a mean frequency of $283.5\pm1.3$~GHz and a loss of $3.18\pm0.13$~dB/mm. These extracted values are also included in Supplementary Fig.~\ref{FigS_C2}b and in Fig.~2e of the main text as purple crosses.

In addition, Supplementary Fig.~\ref{FigS_C2}c shows the tunability of the fundamental resonance frequency with CPS-cavity length $L_\mathrm{cps}$, demonstrating that the CPS resonance can be engineered to satisfy the commensurability condition with the optical resonances.

\newpage
\section{Representative spectra of frequency generation-detection up to 2.5~THz}\label{SI_Spectra}

\begin{figure}[h]
    \centering
    \includegraphics[width=1\textwidth]{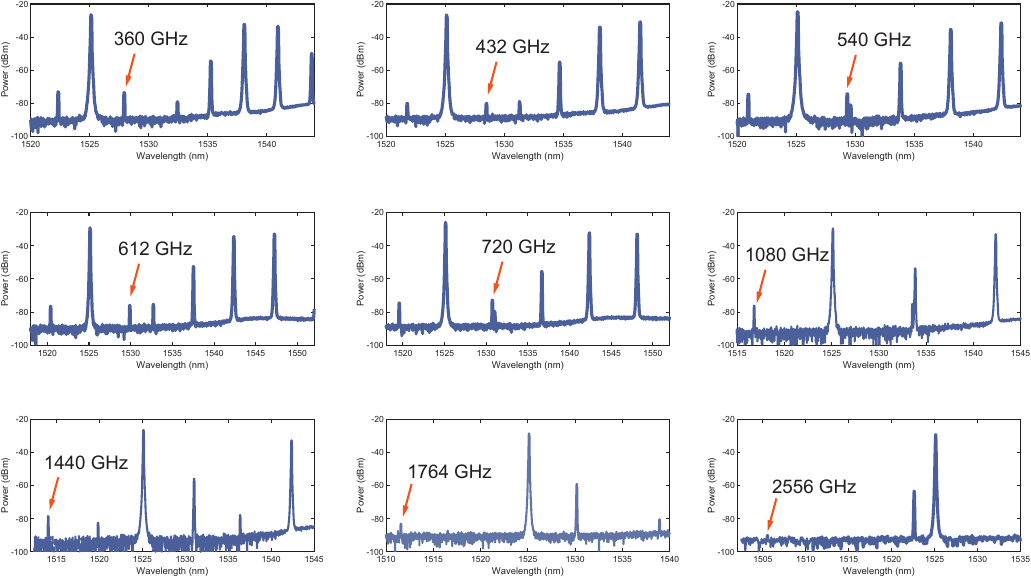}
    \caption{\textbf{Representative spectra of frequency generation-detection.} 
    }
    \label{FigS_D}
\end{figure}

\newpage
\section{Estimation of the CPS–CPS coupling coefficient $\kappa$}\label{SI_CST_split}

\begin{figure}[h!]
    \centering
    \includegraphics[width=1\textwidth]{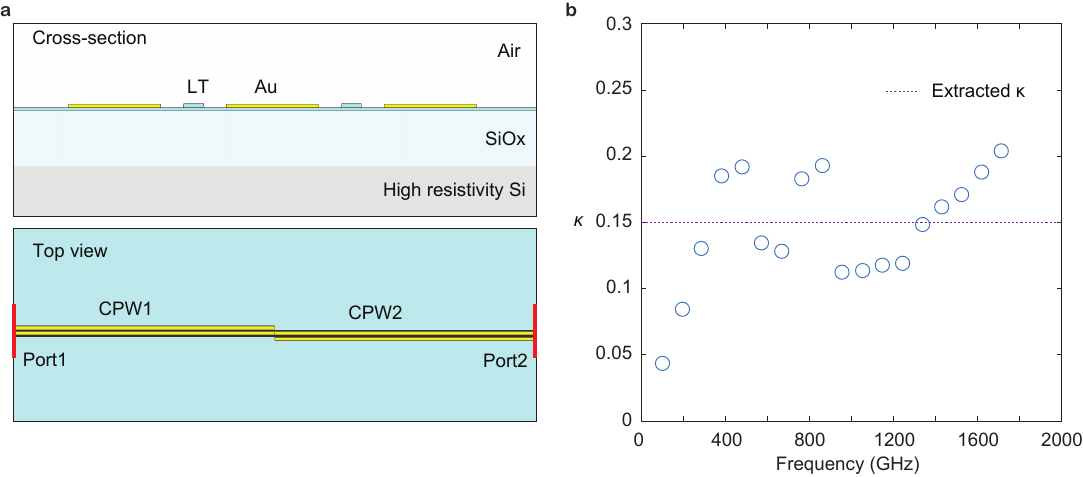}
    \renewcommand{\baselinestretch}{1} 
    \caption{
    \textbf{Numerical extraction of the CPS–CPS coupling coefficient.}
    \textbf{a}. Two-port simulation geometry in CST Studio Suite. The coplanar-stripline (CPS) modes are launched at Port 1, and the outgoing field at Port 2 is projected onto the corresponding CPS modes.
    All material models are lossless, and the total simulated CPS length is reduced to 100~\textmu m.
    \textbf{b}. Frequency-dependent amplitude coupling coefficient $\kappa=|S_{21}|$ extracted from the shortened, lossless coupled-CPS structure. The circles show the simulated values of $S_{21}$, and the purple horizontal line marks $\kappa=0.15$, the value used in the coupled-mode calculations.
    }
    \label{FigS_E}
\end{figure}

We estimated the dimensionless amplitude-coupling coefficient between the two CPS structures using an independent two-port electromagnetic simulation in CST Studio Suite, as illustrated in Supplementary Fig.~\ref{FigS_E}a. 
The modes of the first CPS was launched at Port 1, and the outgoing field at Port 2 was projected onto the corresponding CPS mode.
The resulting complex scattering parameter $S_{21}$ is the ratio of the outgoing modal amplitude at Port 2 to the incident modal amplitude at Port 1.
The corresponding amplitude coupling coefficient is $\kappa=|S_{21}|$.

We made two modifications to reduce the computational complexity and isolate the non-dissipative coupling between the CPS structures.
First, all material models are made lossless, thereby eliminating dissipative contributions such as ohmic loss in the gold electrodes and dielectric absorption. Second, the total simulated CPS length is reduced to 100~\textmu m to decrease the computational cost. The transverse geometry, CPS separation, and longitudinal extent of the coupling region were otherwise kept unchanged.

Supplementary Fig.~\ref{FigS_E}b summarizes the extracted frequency-dependent coupling coefficient $\kappa$. The mean value is $\kappa\approx0.15$, as indicated by the purple horizontal line. 
The departure from this value at the low-frequency end of the simulated range is attributed to the cut-on behavior of the capacitive coupling structure.
The larger, spectrally narrow deviations near 400 and 800 GHz are consistent with standing-wave resonances in the shortened CPS sections or with the excitation of substrate modes.

\newpage

\section{Extraction of the single-photon electro-optic coupling rate $g_0$}\label{SI_g0}

\begin{figure}[h!]
    \centering
    \includegraphics[width=\textwidth]{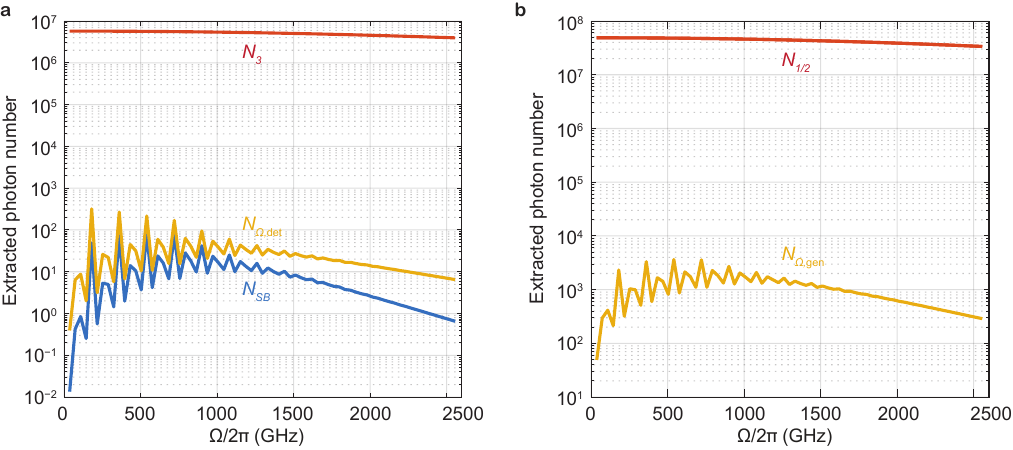}
    \renewcommand{\baselinestretch}{1} 
    \caption{
    \textbf{Photon numbers of the modes involved in (a) anti-Stokes sideband upconversion for mmWave–THz detection and (b) downconversion for mmWave-THz generation.}
    }
    \label{FigS_F}
\end{figure}

We extract the single-photon electro-optic coupling rate $g_0$ using a three-wave-mixing model.
The interacting modes are the optical pump mode $\hat{a}_3$, the optical anti-Stokes sideband mode $\hat{a}_\mathrm{SB}$, and the mmWave-THz mode $\hat{b}$, with $\omega_\mathrm{SB}=\omega_\mathrm{3}+\Omega$. 
Under the rotating-wave approximation, the interaction Hamiltonian is
\begin{equation}
\hat{H}_\mathrm{int}=\hbar g_0
\left(
\hat{a}_\mathrm{SB}^\dagger \hat{a}_3 \hat{b}
+\hat{a}_\mathrm{SB} \hat{a}_3^\dagger \hat{b}^\dagger
\right),
\end{equation}
where $g_0$ is an angular-frequency coupling rate. The equation of motion for the sideband mode is
\begin{equation}
\dot{\hat{a}}_\mathrm{SB}=\left(i\Delta_\mathrm{SB}-\frac{\kappa_\mathrm{SB}}{2}\right)\hat{a}_\mathrm{SB}-i g_0 \hat{a}_3 \hat{b}.
\end{equation}
where $\kappa_\mathrm{SB}$ is the total energy-decay rate of the optical sideband mode expressed in angular-frequency units, and $\Delta_\mathrm{SB}=\omega_\mathrm{3}+\Omega-\omega_\mathrm{SB,cav}$ is the detuning of the generated anti-Stokes sideband from the optical cavity resonance. In our measurements $\kappa_\mathrm{SB}=2\pi \cdot160$~MHz.

In the weak-conversion regime, pump depletion and conversion-induced dynamical backaction are negligible. We therefore replace the coherently driven pump and mmWave-THz operators by their mean amplitudes
\begin{equation}
\hat{a}_3 \rightarrow \alpha_3,
\qquad
\hat{b} \rightarrow \beta_\Omega,
\end{equation}
The corresponding mean coherent intracavity photon numbers are
\begin{equation}
N_3 = |\alpha_3|^2,
\qquad
N_\mathrm{\Omega,det} = |\beta_\Omega|^2.
\end{equation}

The steady-state sideband amplitude is
\begin{equation}
\alpha_\mathrm{SB}=\frac{-i g_0 \alpha_3 \beta_\Omega}{\kappa_\mathrm{SB}/2-i\Delta_\mathrm{SB}}.
\end{equation}
The mean coherent photon number in the sideband mode is therefore
\begin{equation}
N_\mathrm{SB}=|\alpha_\mathrm{SB}|^2=\frac{g_0^2 N_3 N_\Omega}{(\kappa_\mathrm{SB}/2)^2+\Delta_\mathrm{SB}^2}.
\end{equation}
Solving for the single-photon coupling rate gives
\begin{equation}\label{Eq_g0}
g_0=\sqrt{\frac{N_\mathrm{SB}\left[(\kappa_\mathrm{SB}/2)^2+\Delta_\mathrm{SB}^2\right]}{N_3 N_\Omega}}.
\end{equation}

We next relate the photon numbers in these modes to the on-chip intracavity powers inferred from our measurements. 
For each mode $j\in\{\mathrm{SB},3,\Omega\}$, the corresponding modal power is
\begin{equation}\label{Eq_ModePowers}
P_j=\frac{1}{2}\varepsilon_0 c n_j |A_j|^2 S_j
\end{equation}
where $A_j$, $n_j$, and $S_j$ are the electric-field amplitude, refractive index, and effective mode area of mode $j$, respectively.
The energy stored in mode $j$ is obtained from the modal power as
\begin{equation}
U_j=P_j\frac{L_{\mathrm{eff},j}}{v_{g,j}}
\end{equation}
where $v_{g,j}=c/n_{g,j}$ is the group velocity. In the traveling-wave configuration, in which the mmWave–THz field co-propagates with the optical probe,
\begin{equation}
L_{\mathrm{eff},\Omega}=L_\mathrm{cps}
\end{equation}

The photon number in each mode is then
\begin{equation}\label{Eq_photonNumber}
N_j=\frac{P_j n_{g,j}L_{\mathrm{eff},j}}
{\hbar\omega_j c}.
\end{equation}

Using this model and the experimental power values reported in  Supplementary Information~\ref{SI_theory}.7, we extract the intracavity photon numbers in the detector module as shown in Supplementary Fig.~\ref{FigS_F}a. 
For example, at $\Omega/2 \pi=360$~GHz, the average sideband power reported in Supplementary Fig.~\ref{FigS_A7} is $P_\mathrm{SB}=75.3$~pW.
Using Eq.~(\ref{Eq_photonNumber}), we obtain $N_\mathrm{SB}=38.6$, $N_\mathrm{\Omega,det}=241$, and $N_3\approx5.7\times10^6$. Substituting these values into Eq.~(\ref{Eq_g0}) yields $g_0/2\pi\approx13.4$~kHz. 

The preceding derivation applies to mmWave-to-optical upconversion in the detection process. By reciprocity, the downconversion process for generating a mmWave–THz field from two optical pumps can be described analogously.
In this case, the lower-frequency optical mode $\hat{a}_1$ and the higher-frequency optical mode $\hat{a}_2$ are separated by the mmWave-THz frequency $\Omega$, such that $\omega_2-\omega_1=\Omega$. 
The interaction creates a mmWave–THz photon while converting a photon from the higher-frequency optical mode into the lower-frequency optical mode.

The equation of motion for the mmWave-THz mode is
\begin{equation}
\dot{\hat{b}}=
\left(
i\Delta_\Omega-\frac{\kappa_\Omega}{2}
\right)\hat{b}
-i g_0 \hat{a}_1^\dagger \hat{a}_2 ,
\end{equation}
where $\kappa_\Omega$ is the total energy-decay rate of the mmWave-THz mode, expressed in angular-frequency units, and
\begin{equation}
\Delta_\Omega=\omega_2-\omega_1-\Omega_{\mathrm{cav}}
\end{equation}
is the detuning of the optical difference frequency from the mmWave-THz cavity resonance.

In the weak-conversion regime, depletion of the two optical pumps and conversion-induced backaction can again be neglected. We therefore replace the two coherently driven optical operators with their mean amplitudes:
\begin{equation}
\hat{a}_2\rightarrow\alpha_2,
\qquad
\hat{a}_1\rightarrow\alpha_1 .
\end{equation}
The corresponding coherent intracavity photon numbers are
\begin{equation}
N_2=|\alpha_2|^2,
\qquad
N_1=|\alpha_1|^2.
\end{equation}

The steady-state amplitude of the generated mmWave-THz field is then
\begin{equation}
\beta_\Omega=
\frac{-i g_0 \alpha_1^*\alpha_2}
{\kappa_\Omega/2-i\Delta_\Omega}.
\end{equation}
Consequently, the mean coherent photon number in the generated mmWave–THz mode is
\begin{equation}
N_\mathrm{\Omega,gen}=|\beta_\Omega|^2=
\frac{g_0^2 N_1 N_2}
{(\kappa_\Omega/2)^2+\Delta_\Omega^2}.
\end{equation}
Solving for the single-photon coupling rate gives
\begin{equation}\label{Eq_g0_DFG}
g_0=\sqrt{
\frac{N_\mathrm{\Omega,gen}\left[(\kappa_\Omega/2)^2+\Delta_\Omega^2\right]}
{N_1 N_2}
}.
\end{equation}

Using this model and the experimental power values reported in  Supplementary Information~\ref{SI_theory}.7, we extract the intracavity photon numbers as shown in Supplementary Fig.~\ref{FigS_F}b. 
In this case, we obtain $N_1=N_2=4.85\times10^7$, and $N_\mathrm{\Omega,gen}\approx3323$. Substituting these values into Eq.~(\ref{Eq_g0_DFG}) yields $g_0/2\pi\approx11.9$~kHz.
Here, $\kappa_\Omega\approx2\pi\cdot20$~GHz is estimated from direct measurement of a neighbor mode at 288~GHz, shown in Supplementary Fig.~\ref{FigS_C}f, and from the numerical simulations shown in Supplementary Fig.~\ref{FigS_B}e.

Therefore, the values of $g_0/2\pi$ independently extracted from the two processes agree reasonably well, providing mutual validation.
We then apply the upconversion analysis to all frequencies measured in Supplementary Fig.~\ref{FigS_F}a. 
For frequencies detuned from the CPS resonances, such as 108 and 144 GHz, we must also account for the mmWave–THz detuning when calculating the intracavity mmWave–THz photon number $N_\mathrm{\Omega,det}$. In general, the coherent intracavity amplitude $b$ of a driven microwave mode (the detector CPS mode in our experiment) obeys
\begin{equation}
\dot{b}=\left(i\Delta_{\Omega}-\frac{\kappa_{\Omega}}{2}\right)b
+\sqrt{\kappa_{\Omega,\mathrm{in}}}\ b_{\mathrm{in}},
\label{eq:si_microwave_eom}
\end{equation}
where $\kappa_{\Omega}$ is the total microwave energy-decay rate, $\kappa_{\Omega,\mathrm{in}}$ is the coupling rate of the driven microwave port, and $\Delta_\Omega$ is the detuning of the microwave drive from the CPS resonance. 
Writing the steady-state intracavity amplitude as $\beta_\Omega$, we obtain
\begin{equation}
\beta_{\Omega}=\frac{\sqrt{\kappa_{\Omega,\mathrm{in}}} b_{\mathrm{in}} }
{\kappa_{\Omega}/2-i\Delta_{\Omega}}.
\label{eq:si_microwave_amplitude}
\end{equation}
For an input microwave power $P_{\Omega,\mathrm{in}}$, the photon-flux-normalized input field satisfies
\begin{equation}
\left|b_{\mathrm{in}}\right|^{2}=\frac{P_{\Omega,\mathrm{in}}}
{\hbar\Omega}.
\end{equation}
The corresponding intracavity microwave photon number is therefore
\begin{equation}
N_{\Omega}(\Delta_{\Omega})=\left|\beta_{\Omega}\right|^{2}
=\frac{\kappa_{\Omega,\mathrm{in}} }
{\left(\kappa_{\Omega}/2\right)^{2}+ \Delta_{\Omega}^{2}}
\frac{ P_{\Omega,\mathrm{in}} }
{\hbar\Omega}.
\label{eq:si_detuned_microwave_occupation}
\end{equation}

We define $N_{\Omega}^{(0)}$ as the microwave photon number produced by the same incident power when the microwave drive is resonant with the CPS cavity. In this case,
\begin{equation}
N_{\Omega}^{(0)}=\frac{\kappa_{\Omega,\mathrm{in}}}
{\left(\kappa_{\Omega}/2\right)^{2}}
\frac{P_{\Omega,\mathrm{in}}}
{\hbar\Omega}.
\label{eq:si_resonant_microwave_occupation}
\end{equation}
The detuned and resonant photon numbers are therefore related by
\begin{equation}
N_{\Omega}(\Delta_{\Omega})=\frac{N_{\Omega}^{(0)}}
{1+\left(2\Delta_{\Omega}/\kappa_{\Omega}\right)^{2}}.
\label{eq:si_microwave_lorentzian}
\end{equation}
Substituting Eq.~\eqref{eq:si_microwave_lorentzian} into Eq.~\eqref{Eq_g0} gives
\begin{equation}
g_{0}=\sqrt{\frac{N_\mathrm{SB}\left[
\left(\kappa_\mathrm{SB}/2\right)^{2}+\Delta_{\mathrm{SB}}^{2}
\right]}
{N_{3}N_\mathrm{\Omega,det}^{(0)}}
}
\sqrt{1+\left(\frac{2\Delta_{\Omega}}
{\kappa_{\Omega}}
\right)^{2}
}
\label{eq:si_g0_resonance_referenced}
\end{equation}

Using this expression, we extract the values of $g_0/2\pi$ shown by the blue dots in Fig. 3f of the main text.

\newpage

\section{Dual-pump loading in lithium tantalate and lithium niobate}\label{SI_DualPump}

\begin{figure}[p!]
    \centering
    \includegraphics[width=1\textwidth]{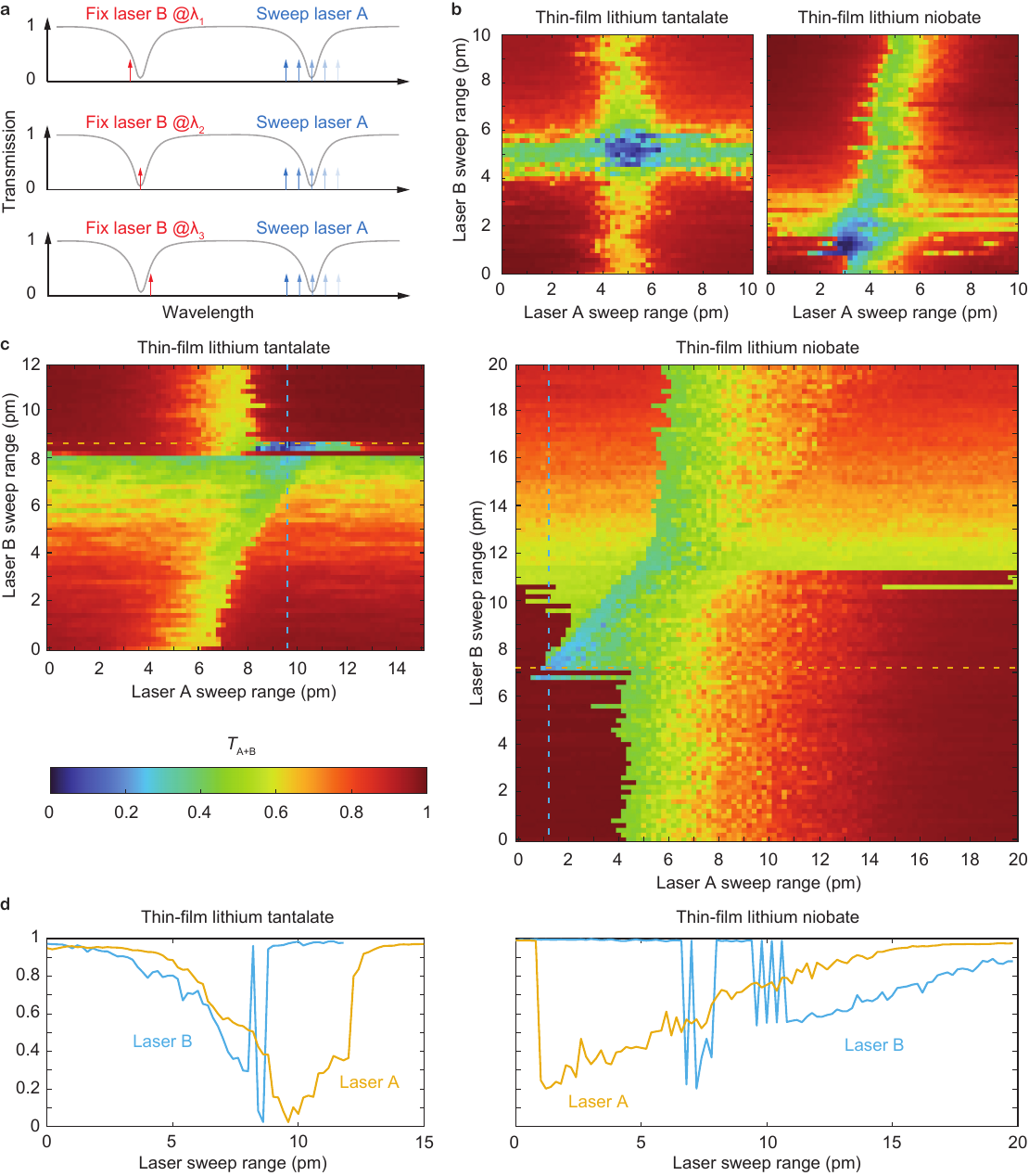}
    \renewcommand{\baselinestretch}{1} 
    \caption{
    \textbf{Dual-pump loading in matched lithium tantalate and lithium niobate resonators.}
    \textbf{a}. Dual-laser wavelength-sweep protocol. Lasers A and B address two adjacent optical resonances. Laser B is stepped in the outer loop, whereas laser A is swept across its resonance in the inner loop.
    The power-normalized combined through-port transmission is denoted by $T_\mathrm{A+B}$, from which the operational loading metric is defined as $\eta_\mathrm{load}  = 1-T_\mathrm{A+B}$. 
    \textbf{b}. Two-dimensional maps of $T_\mathrm{A+B}$ for matched lithium tantalate (left) and lithium niobate (right) resonators at a total on-chip pump power of 0.12~mW, corresponding to an intracavity power of 7~mW under full loading. 
    The axes show the wavelength sweeps of lasers A and B from their corresponding cavity resonances. Both devices exhibit minima with $T_\mathrm{A+B}<0.05$, corresponding to a combined loading metric above 95\%.
    \textbf{c}. Dual-pump transmission maps after a tenfold increase in the total off-resonant on-chip pump power to 1.2~mW, corresponding to an intracavity power of approximately 70~mW under full loading. The lithium tantalate resonances shift toward longer wavelengths, whereas the lithium niobate resonances shift toward shorter wavelengths.
    \textbf{d}. Line cuts through the regions of maximum loading in c. The lithium tantalate retains $\eta_\mathrm{load} > 97\%$, while in lithium niobate $\eta_\mathrm{load} = 75\%$.
    }
    \label{FigS_G}
\end{figure}

To compare simultaneous dual-pump loading in thin-film lithium tantalate and lithium niobate, we fabricate the two chips side by side using identical electron-beam lithography, dry-etching, and post-etch redeposition-removal processes.
Both devices are fabricated from 600-nm-thick X-cut films. The fabricated resonators have measured waveguide widths of 2~\textmu m, etch depths of 350~nm, sidewall angles of $75^\circ$, and loaded quality factors of approximately 1 million for critically coupled resonators. 
Together, these measured parameters, rather than the common fabrication process alone, establish how closely the two devices were matched.

Each chip is thermally anchored to a temperature-controlled stage at $297\pm0.01$~K. 
Two lasers, denoted A and B, address adjacent cavity modes.
We acquire a two-dimensional transmission map using nested wavelength sweeps: laser B is stepped in the outer loop, whereas laser A is swept across its resonance in the inner loop, as illustrated in Supplementary Fig.~\ref{FigS_G}a.

We denote the off-resonant transmitted powers of the two lasers by $P_\mathrm{A}^0$ and $P_\mathrm{B}^0$, and their corresponding normalized through-port transmissions by $T_\mathrm{A}$ and $T_\mathrm{B}$. The combined transmission in measurements is
\begin{equation}
T_\mathrm{A+B}= \frac{P_\mathrm{A}^0T_\mathrm{A}+P_\mathrm{B}^0T_\mathrm{B}}{P_\mathrm{A}^0+P_\mathrm{B}^0}
\end{equation}
We then define the operational dual-pump loading metric as
\begin{equation}
\eta_\mathrm{load}  = 1-T_\mathrm{A+B}
\end{equation}
This quantity measures the power-weighted through-port extinction and serves as a proxy for simultaneous cavity loading by the two pumps.

At a total on-chip pump power of 0.12 mW, corresponding to an intracavity power of approximately 7 mW under full loading, the two-dimensional maps exhibit a minimum combined transmission of $T_\mathrm{A+B}<0.05$ for both materials, as shown in Supplementary Fig.~\ref{FigS_G}b). 
The corresponding dual-pump loading metric therefore exceeds 95\% under moderate-power operation.
For clarity, the plotted axes represent the wavelength-sweep ranges rather than absolute wavelengths.

We then increase the total on-chip pump power by a factor of ten, to 1.2 mW, corresponding to an intracavity power of approximately 70 mW under full loading. 
The resulting two-dimensional maps are shown in Supplementary Fig.~\ref{FigS_G}c, and line cuts through the regions of maximum loading are shown in Supplementary Fig.~\ref{FigS_G}d.
As a result, the lithium tantalate resonances shift toward longer wavelength by approximately 5~pm, whereas the lithium niobate resonances shift toward shorter wavelengths by more than 10 pm.
The directions of these resonance shifts are consistent with a predominantly thermo-optic redshift in lithium tantalate and a stronger photorefractive contribution producing a blueshift in lithium niobate, although the shift direction alone does not uniquely separate the two mechanisms~\cite{yu2024tunable}.

In lithium tantalate, the two resonances retained an overlapping low-transmission region with $\eta_\mathrm{load} > 97\%$. 
In lithium niobate, abrupt switching to the low-power resonance branch occurred before the through-port transmissions of both pumps could be minimized simultaneously. 
This switching limited the maximum combined loading metric to approximately $\eta_\mathrm{load} = 75\%$.

\newpage

\section{Cryogenic packaging and measurements}\label{SI_Cryo}

\begin{figure}[h!]
    \centering
    \includegraphics[width=1\textwidth]{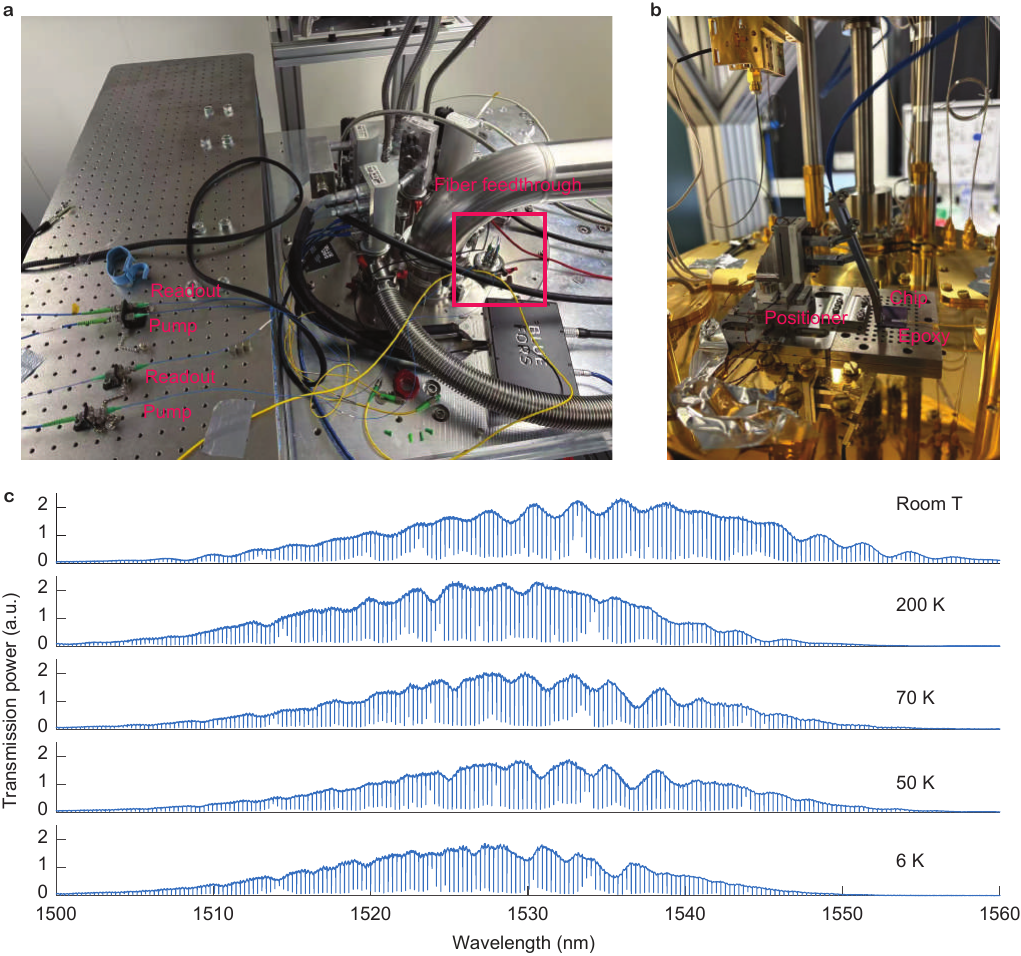}
    \renewcommand{\baselinestretch}{1} 
    \caption{
    \textbf{Cryogenic fiber packaging and optical access.}
    \textbf{a}. Cryogenic configuration used to interface the photonic signal chain. Separate two-channel fiber arrays provide optical input and output for the generator and detector modules. The generator-side array is permanently attached using cryogenic-compatible epoxy, whereas the detector-side array is aligned in situ using a cryogenic piezoelectric positioner. Four fibers enter the dilution refrigerator through two hermetic feedthroughs and are thermally anchored at the 50-K and 4-K plates before reaching the chip assembly.
    \textbf{b}. The system is mounted on the still plate inside the dilution refrigerator, working at 4~K.
    \textbf{c}. Optical transmission spectra of the detector racetrack resonator recorded during cooling from room temperature to approximately 4~K.
    }
    \label{FigS_H1}
\end{figure}

\begin{figure}[h!]
    \centering
    \includegraphics[width=1\textwidth]{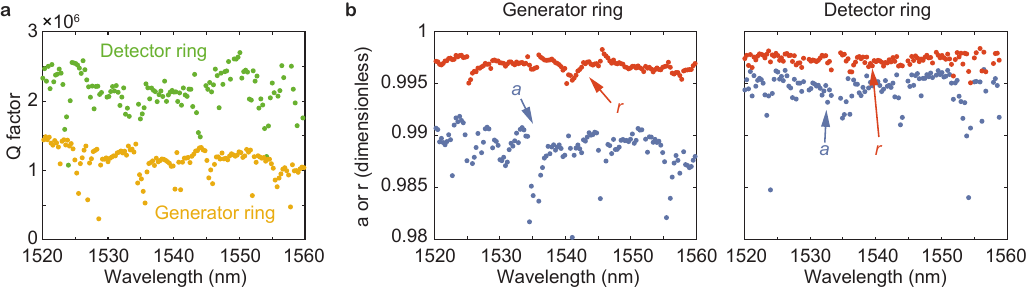}
    \renewcommand{\baselinestretch}{1} 
    \caption{
    \textbf{Optical characterization of the generator and detector racetrack resonators.}
    \textbf{a}. Quality factors (Q) extracted from optical transmission spectra measured over the wavelength range from 1520 to 1560~nm.  
    \textbf{b}. Extracted round-trip field-transmission coefficient $a$ and field-reflection coefficient $r$ for the generator racetrack resonator (left panel) and the detector racetrack resonator (right panel). 
    }
    \label{FigS_H2}
\end{figure}

To characterize cryogenic operation, we interface the generator and detector modules with separate two-channel fiber arrays, each providing one input and one output channel.
The generator fiber array is aligned at room temperature and secured using cryogenic-compatible epoxy, whereas the detector fiber array is mounted on a cryogenic piezoelectric positioner to enable in situ alignment.
Four fibers enter the dilution refrigerator through two hermetic feedthroughs, one of which is shown in Supplementary Fig.~\ref{FigS_H1}a,
The fibers are carefully routed and thermally anchored to the 50~K and 4~K plates.
The system is mounted farther down, on the still plate, within a shielded environment maintained at approximately 4 K, as shown in Supplementary Fig.~\ref{FigS_H1}b. 

We monitor the optical transmission spectra throughout pump-down and cooling, as shown in Supplementary Fig.~\ref{FigS_H1}c. 
The measured transmission changes by less than 1 dB between ambient conditions and the operating temperature of approximately 4 K, while the optical resonances exhibit a slight shift toward shorter wavelengths.
The positioner-mounted detector interface undergoes micrometer-scale lateral drift and is therefore realigned approximately every 3 h using the piezoelectric positioner.
By contrast, the transmission through the epoxy-fixed generator interface remains well-aligned.

Once the system reaches its operating temperature of approximately 4~K, we repeat the end-to-end measurements using the same protocol as that used at room temperature.
Detector sidebands are resolved at frequencies between 180 and 576 GHz, as shown in Supplementary Fig.~\ref{FigS_H3}a. 

The optical pump and probe powers $P_{\omega1}$, $P_{\omega2}$, and $P_{\omega3}$, together with the sideband power $P_\mathrm{SB}$, are measured directly. 
The mmWave–THz powers are inferred using the calibrated generation–transfer–detection model described in the Supplementary Information~\ref{SI_theory}.

In the signal core fabricated for the cryogenic experiments, the generator and detector racetrack resonators have round-trip field-transmission coefficients of
$a_\mathbf{gen}=0.9888\pm0.0003$ and $a_\mathbf{det}=0.9948\pm0.0004$, respectively.
Their corresponding field-reflection coefficients are
$r_\mathbf{gen}=0.9972\pm0.0002$ and $r_\mathbf{det}=0.9970\pm0.0002$, respectively, as shown in Supplementary Fig.~\ref{FigS_H2}. 
Using Eq~(\ref{Eq_FE_gen}), these values give a field-enhancement factor of 5.4, corresponding to a power-enhancement factor of 29 in the generator. 
In the detector resonator, the externally injected optical probe at $\omega_3$ has a field-enhancement factor of 9.6, corresponding to a power-enhancement factor of 92.
For the internally generated sideband field, using Eq.~\ref{Eq_FE_det}), the field-enhancement factor is 121.8 and the corresponding power-enhancement factor is approximately 14800.

\begin{figure}[h!]
    \centering
    \includegraphics[width=1\textwidth]{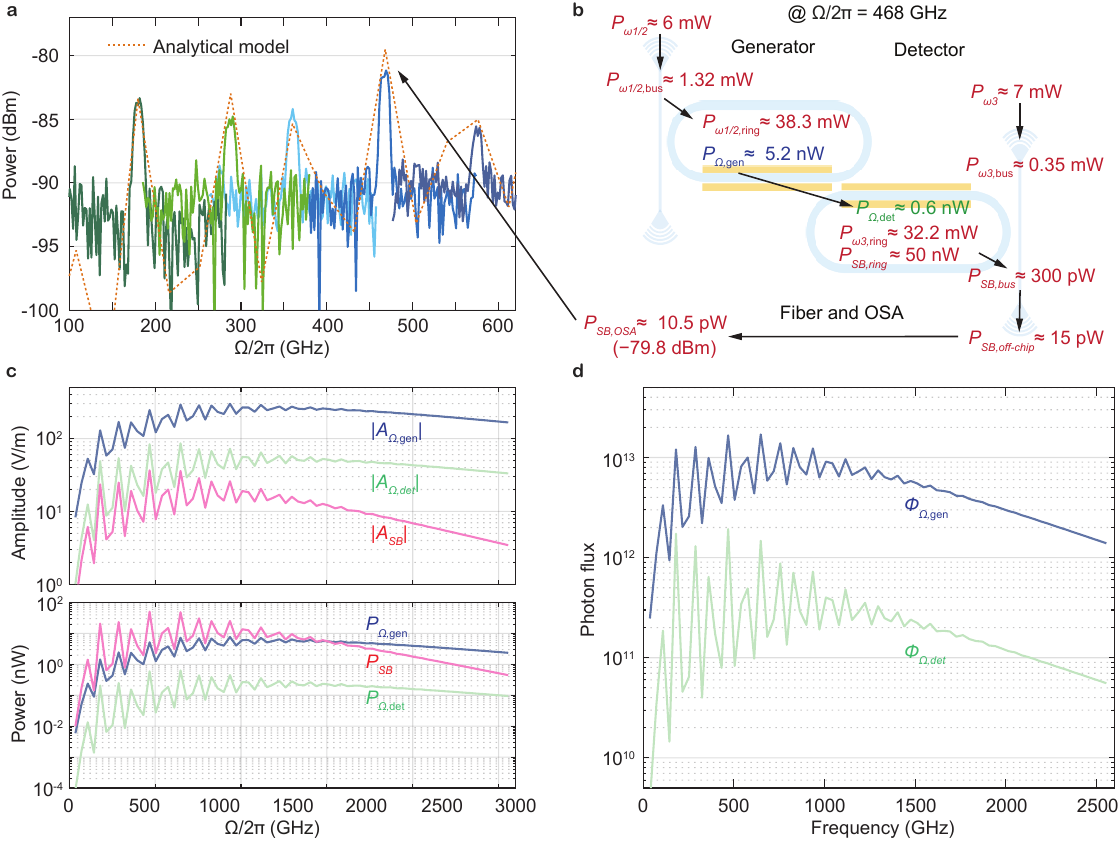}
    \renewcommand{\baselinestretch}{1} 
    \caption{
    \textbf{End-to-end signal core operation under cryogenic conditions.}
    \textbf{a}.Optical sideband powers at the detector output measured at selected frequencies between 180 and 576~GHz at approximately 4~K. The yellow curve shows the corresponding analytical results from the generation–transfer–detection model.
    \textbf{b}. Calibrated on-chip power flow through the generation–transfer–detection chain at 468~GHz. 
    \textbf{c}. Model-inferred on-chip field amplitudes and powers of the generated mmWave–THz field, transferred mmWave–THz field, and detected optical sideband over the measured frequency range, together with analytical extrapolations up to 2.5 THz.
    \textbf{d}. Generated mmWave–THz photon flux calculated for the generator (blue) and detector CPS cavity (green).
    }
    \label{FigS_H3}
\end{figure}

Supplementary Fig.~\ref{FigS_H3}b and Table~\ref{Table_PowerFlow_Bluefors} summarize the on-chip power flow when measured at 468~GHz as an example. 
On the generation side, the two pump lasers each have an output power of approximately 15~mW. Their outputs are combined using a 50:50 fiber coupler, resulting in a total output power of approximately 15~mW across the two optical tones. After accounting for an optical path loss of approximately 20\%, the off-chip pump powers at the device input are $P_{\omega_1}\approx P_{\omega_2}\approx6$~mW. 
With a generator-side grating-coupler efficiency of approximately 22\%, the on-chip bus-waveguide powers are $P_{\omega_\mathrm{1,bus}}\approx P_\mathrm{\omega_2,bus}\approx 1.32$~mW. 
A power-enhancement factor of approximately 29 results in circulating powers of $P_\mathrm{\omega1,ring}\approx P_\mathrm{\omega_2,ring}\approx 38.3$~mW inside the generator ring resonator.

Using Eqs.~(\ref{Eq_gen}) and (\ref{Eq_THzPower}), we obtain a generated mmWave–THz power of approximately $P_\mathrm{\Omega,gen}=5.2$~nW in the generator CPS cavity, and a transferred power of $P_{\Omega,\mathrm{det}} \approx0.6$~nW in the detector CPS cavity calculated via Eqs.~(\ref{Eq_det}) and (\ref{Eq_THzPower}). 

On the detection side, the racetrack resonator is probed using a third laser with an output power of 10~mW. After accounting for an optical path loss of approximately 30\%, the off-chip probe power at the device input is $P_{\omega_1}\approx P_{\omega_2}\approx7$~mW. 
The grating-coupler efficiency on the detector side is approximately 5\%, resulting in an on-chip probe power of $P_{\omega_\mathrm{3,bus}}\approx 0.35$~mW. 
With a power-enhancement factor of approximately 92, the calculated circulating probe power inside the detector racetrack resonator is $P_{\omega_\mathrm{3,ring}}\approx32.2$~mW.

Using Eqs.~(\ref{Eq_A_SB}) and (\ref{Eq_P_SB}), the calculated circulating power of the upconverted sideband in the detector racetrack resonator is $P_\mathrm{SB,ring}\approx50$~nW, including the power-enhancement factor of approximately 14800 for an internally generated field. 
The sideband power coupled into the bus waveguide is $P_\mathrm{SB,bus}=P_\mathrm{SB,ring}\cdot(1-r^2)$, where 
$r\approx0.997$ is the detector field-reflection coefficient.
Therefore, we obtain $P_\mathrm{SB,bus}\approx300$~pW. 
After accounting for the detector-side grating-coupler efficiency of 5\%, the off-chip sideband power is $P_\mathrm{SB,off-chip}\approx15$~pW.
An additional insertion loss of approximately 30\% before the optical spectrum analyzer results in a measured sideband power of $P_{\mathrm{SB,OSA}}\approx10.5$~pW, consistent with Supplementary Fig.~\ref{FigS_H3}a.

\begin{table}[h]
\centering
\caption{\textbf{Power flow on chip operated at 468~GHz.}}
\label{Table_PowerFlow_Bluefors}
\begin{tabular}{lll}
\hline
Parameter & Symbol & Value \\
\hline
Laser pump power (off-chip) & $P_{\omega1}$ and $P_{\omega2}$ & $6~\mathrm{mW}$ \\
Laser pump power (in bus waveguide) & $P_\mathrm{\omega1,bus}$ and $P_\mathrm{\omega2,bus}$ & $1.32~\mathrm{mW}$ \\
Intracavity pump power of generator ring & $P_{\omega1,ring}$ and $P_{\omega2,ring}$ & $38.3~\mathrm{mW}$ \\
Generated mmWave-THz power in generator CPS & $P_\mathrm{\Omega,gen}$ & $90~\mathrm{nW}$ \\
Transferred mmWave-THz power in detector CPS & $P_\mathrm{\Omega,det}$ & $2.9~\mathrm{nW}$ \\
Laser probe power & $P_{\omega3}$ & $3~\mathrm{mW}$ \\
Intracavity probe power of detector ring & $P_\mathrm{\omega3,ring}$ & $12~\mathrm{mW}$ \\
Intracavity sideband power of detector ring & $P_\mathrm{SB,ring}$ & $55~\mathrm{nW}$ \\
Off-chip sideband power & $P_\mathrm{SB,off-chip}$ & $45~\mathrm{pW}$ \\
Measured sideband power in OSA & $P_\mathrm{SB,OSA}$ & $15~\mathrm{pW}$ \\
\hline
\end{tabular}
\end{table}

Supplementary fig.~\ref{FigS_H3}c summarizes the model-inferred on-chip electric-field amplitudes and powers over the measured cryogenic frequency range.
The analytical model is also used to extrapolate these quantities over the full frequency range up to 2.5~THz. 
We convert the inferred mmWave–THz powers into photon fluxes using $\Phi_\mathbf{\Omega,gen}=\frac{P_\mathrm{\Omega,gen}}{\hbar \Omega}$ for the generator CPS cavity and $\Phi_\mathbf{\Omega,det}=\frac{P_\mathrm{\Omega,det}}{\hbar \Omega}$ for the detector CPS cavity. The resulting fluxes reach approximately $\Phi_\mathbf{\Omega,gen}\approx10^{13}$ photons per second and $\Phi_\mathbf{\Omega,det}\approx10^{12}$ photons per second, as shown in Supplementary Fig.~\ref{FigS_H3}d.

For an order-of-magnitude comparison, consider a resonator with a mean photon occupation of $\bar{n}=10$ and an angular energy-decay rate $\gamma_\Omega$. The corresponding photon-loss flux is
$\Phi_\mathbf{\Omega,loss}=\bar{n}\cdot \gamma_\Omega$. 
For resonator linewidths ranging from $\gamma_\Omega/2\pi=$ 10~MHz to 1~GHz, the photon-loss flux ranges from approximately $\Phi_\mathbf{\Omega}\approx 6.28\times 10^{8}$ to $6.28\times 10^{10}$ photons per second, as shown in Fig.~5e of the main text.

Based on the model scaling, we estimate that these photon fluxes can be reached with on-chip powers in each generation pump of approximately $P_{\omega_\mathrm{1/2,bus}}\approx10$ and 100~\textmu W, respectively.

Sidebands above 576 GHz are not resolved under these conditions because the generator is operated at a lower coupled pump power than in the room-temperature measurements to limit optical loading of the cryogenic stage.
Nevertheless, the generated and detected signals span a frequency range relevant to mmWave quantum measurements and provide inferred photon fluxes compatible with superconducting-resonator readout.

\newpage
\bibliography{SI_Ref}
\bibliographystyle{paper}